\documentstyle{mn}
\input{psfig}
\newcommand{\etal}{{et al.~}}

\newcommand{\kmsmpc}{\>{\rm km}\,{\rm s}^{-1}\,{\rm Mpc}^{-1}}
\newcommand{\kms}{\>{\rm km}\,{\rm s}^{-1}}

\newcommand{\Mpc}{\>{\rm Mpc}}

\newcommand{\Msun}{\>{\rm M_{\odot}}}
\newcommand{\Lsun}{\>{\rm L_{\odot}}}
\newcommand{\MLsun}{\>({\rm M}/{\rm L})_{\odot}}

\newcommand{\walpha}{\tilde{\alpha}}
\newcommand{\wLstar}{\tilde{L}^{*}}

\newcommand{\beq}{\begin{equation}}
\newcommand{\eeq}{\end{equation}}

\newcommand{\msunh}{\>h^{-1}\rm M_\odot}

\def\gtsima{$\; \buildrel > \over \sim \;$}
\def\ltsima{$\; \buildrel < \over \sim \;$}
\def\prosima{$\; \buildrel \propto \over \sim \;$}
\def\gsim{\lower.7ex\hbox{\gtsima}}
\def\lsim{\lower.7ex\hbox{\ltsima}}
\def\simgt{\lower.7ex\hbox{\gtsima}}
\def\simlt{\lower.7ex\hbox{\ltsima}}
\def\simpr{\lower.7ex\hbox{\prosima}}
\def\la{\lsim}
\def\ga{\gsim}
\def\lta{\la}
\def\gta{\ga}

\newcommand{\apj}{ApJ}
\newcommand{\apjs}{ApJS}
\newcommand{\aj}{AJ}
\newcommand{\mnras}{MNRAS}

\newdimen\hssize
\newdimen\hdsize
\begin{document}


\title[The Abundance and Radial Distribution of Satellite Galaxies]
      {The Abundance and Radial Distribution of Satellite Galaxies}
\author[van den Bosch, Yang, Mo \& Norberg]
       {Frank C. van den Bosch$^{1}$, 
        Xiaohu Yang$^{2}$, H.J. Mo$^{2}$, and Peder Norberg$^{1}$
        \thanks{E-mail: vdbosch@physics.ethz.ch}\\
        $^1$Department of Physics, Swiss Federal Institute of
         Technology, ETH H\"onggerberg, CH-8093, Zurich,
         Switzerland\\ 
        $^2$Department of Astronomy, University of Massachussets, 710
         North Pleasant Street, Amherst MA 01003-9305, USA}


\date{}

\pagerange{\pageref{firstpage}--\pageref{lastpage}}
\pubyear{2000}

\maketitle

\label{firstpage}


\begin{abstract}
  Using detailed  mock galaxy redshift surveys  (MGRSs) we investigate
  the abundance  and radial  distribution of satellite  galaxies.  The
  mock surveys  are constructed using large  numerical simulations and
  the conditional luminosity function  (CLF), and are compared against
  data from the Two Degree  Field Galaxy Redshift Survey (2dFGRS).  We
  use  Monte  Carlo  Markov  Chains  to  explore  the  full  posterior
  distribution of the  CLF parameter space, and show  that the average
  relation  between  light and  mass  is  tightly  constrained and  in
  excellent agreement with  our previous models and with  that of Vale
  \& Ostriker.   The radial  number density distribution  of satellite
  galaxies in the 2dFGRS reveals a pronounced absence of satellites at
  small projected  separations from their host galaxies.   This is (at
  least partially) due to the  overlap and merging of galaxy images in
  the   APM   catalogue.    Because   of  the   resulting   close-pair
  incompleteness  we  are   unfortunately  unable  to  put  meaningful
  constraints on  the radial  distribution of satellite  galaxies; the
  data are  consistent with a radial number  density distribution that
  follows that of  the dark matter particles, but we  can not rule out
  alternatives  with a  constant number  density  core.  Marginalizing
  over the  full CLF parameter space,  we show that  in a $\Lambda$CDM
  concordance cosmology the observed  abundances of host and satellite
  galaxies in  the 2dFGRS indicate  a power spectrum  normalization of
  $\sigma_8 \simeq 0.7$. The same cosmology but with $\sigma_8=0.9$ is
  unable to  simultanously match the abundances of  host and satellite
  galaxies.   This  confirms our  previous  conclusions  based on  the
  pairwise  peculiar velocity dispersions  and the  group multipliticy
  function.
\end{abstract}




\section{Introduction}
\label{sec:intro}

In  the  hierarchical   formation  scenario,  satellite  galaxies  are
associated with dark  matter subhaloes, which, some time  in the past,
were accreted by  their current host (or parent)  halo.  As satellites
orbit their  host galaxies they  are subjected to various  forces that
try  to dissolve  them:  dynamical friction,  tides  from the  central
object(s) and  impulsive collisions with other  satellites. A detailed
understanding of  the abundance  and radial distribution  of satellite
systems, therefore,  provides important constraints on  the outcome of
these various physical processes  which are an essential ingredient of
galaxy formation.

Traditionally, satellite  galaxies have mainly been  used as kinematic
tracers of  the dark matter potential  well of the  parent halo (e.g.,
Carlberg  \etal  1996; Little  \&  Tremaine  1987;  Evans \etal  2000;
Zaritsky  \etal  1993,  1997;  McKay  \etal 2002;  Prada  \etal  2003;
Brainerd \& Specian 2003; van  den Bosch \etal 2004b). Since satellite
galaxies are distributed over the entire halo, they are ideally suited
to measure  the total virial mass,  more so than  for example rotation
curves, which only probe the potential out to a fraction of the virial
radius.  The  {\it distribution}  of satellite galaxies,  however, has
received considerably  less attention.  In  addition to a  few studies
aimed  at testing  the claim  by  Holmberg (1969)  that the  azimuthal
distribution of satellite galaxies  is anisotropic with respect to the
orientation  of  the  host   galaxy,  the  so-called  Holmberg  effect
(Zaritsky  \etal 1997;  Zaritsky  \& Gonzales  1999;  Sales \&  Lambas
2004),  relatively  few studies  have  focussed  on  the {\it  radial}
distribution  of satellite galaxies  (but see  Lake \&  Tremaine 1980;
Vader \& Sandage 1991; Lorrimer \etal 1994; Willman \etal 2004).

The   interest   in   satellite   galaxies  has   recently   increased
considerably, largely due to  the dramatic increase in computing power
that  has  made  it  possible  to resolve  dark  matter  subhaloes  in
cosmological  numerical simulations (Tormen  1997; Ghigna  \etal 1998;
Klypin \etal 1999; Moore \etal 1999; Stoehr \etal 2002; Kravtsov \etal
2003; De Lucia  \etal 2004; Diemand, Moore \&  Stadel 2004; Gill \etal
2004a,b; Weller,  Ostriker \& Bode  2004; Reed \etal 2004).   This has
resulted in  a number of detailed  studies of the  abundances and both
the  spatial  and  the  velocity  distribution  of  subhaloes.   Since
satellite galaxies are thought  to be associated with these subhaloes,
this has opened the  possibility to compare the statistical properties
of satellite  galaxies directly with  those of dark  matter subhaloes,
resulting in two apparent inconsistencies.

First  of all,  the radial  distribution of  dark matter  subhaloes is
found to  be spatially  anti-biased with respect  to the dark  matter. 
However,  the  observed  distribution  of cluster  galaxies  seems  to
accurately  follow a  Navarro, Frenk  \& White  (1997,  hereafter NFW)
distribution (Beers  \& Tonry 1986;  Carlberg, Yee \&  Ellingson 1997;
van der Marel \etal 2000;  Lin, Mohr \& Stanford 2004).  Although this
might  signal   an  inconsistency  of  the   standard  CDM  framework,
semi-analytical models of galaxy  formation that take the evolution of
dark  matter  subhaloes  into   account  can  reproduce  the  observed
distribution of cluster galaxies  (Springel \etal 2001; Diaferio \etal
2001;  Gao  \etal 2004b).   The  reason  is  that galaxies  are  dense
concentrations of baryons in  the centers of dark matter (sub)-haloes,
which are more resilient to tidal disruption than (the outer parts of)
their dark matter haloes.

The second apparent inconsistency concerns the abundances of satellite
galaxies.  Although the subhalo  mass function, when normalized to the
mass of the parent halo, is  found to be virtually independent of halo
mass  (Moore \etal  1999; De  Lucia  \etal 2004;  Diemand \etal  2004;
Weller \etal 2004; but see Gao \etal 2004a), the satellite populations
of  galaxy clusters  are very  different  from those  of galaxy  sized
haloes (Klypin \etal  1999; Moore \etal 1999; D'onghia  \& Lake 2003). 
This  must be  telling us  something  important about  the physics  of
galaxy  formation,  and  numerous  studies have  focussed  on  various
mechanism to explain this  discrepancy between the number of predicted
subhaloes   and  observed  satellites   (e.g.,  Kauffmann,   White  \&
Guiderdoni  1993; Bullock,  Kravtsov  \& Weinberg  2000; Benson  \etal
2002)

At this point in time there  is a strong need for better observational
constraints   regarding  the   statistical  properties   of  satellite
galaxies.  In  particular, does  the radial distribution  of satellite
galaxies in galaxy sized haloes  follow a NFW profile as for clusters,
or does  it reveal a constant  density core, as found  for dark matter
subhaloes?   How does the  abundance of  satellite galaxies  depend on
halo  mass,  or  on  the  luminosity  of the  host  galaxy?   Is  this
mass-dependent  abundance consistent with  the full  galaxy luminosity
function  and  the halo  mass  function?   The conditional  luminosity
function (hereafter  CLF) formalism developed  by Yang, Mo \&  van den
Bosch (2003) and  van den Bosch, Yang \& Mo  (2003a) is ideally suited
to address  these questions. It  describes how many galaxies  of given
luminosity reside in a halo of  given mass. In this paper we therefore
use  the   CLF,  constrained  using  the   abundances  and  clustering
properties of galaxies in  the Two-Degree Field Galaxy Redshift Survey
(2dFGRS,  Colless \etal  2001),  to study  the  abundances and  radial
distribution of satellite galaxies.  We construct detailed mock galaxy
redshift surveys,  based on  the CLF, for  direct comparison  with the
2dFGRS data.  We  show that a close-pair incompleteness  in the 2dFGRS
prevents us from putting  significant constraints on the radial number
density  distribution of  satellite  galaxies, and  that matching  the
abundances of  host and  satellite galaxies in  the 2dFGRS  requires a
$\Lambda$CDM concordance cosmology with  a relatively low value of the
power-spectrum normalization parameter $\sigma_8$.

This  paper  is organized  as  follows.   In Section~\ref{sec:clf}  we
describe the CLF,  and use a Monte-Carlo Markov  Chain to fully sample
the posterior  distribution of the CLF.   In Section~\ref{sec:mock} we
describe how to use the CLF to construct detailed Mock Galaxy Redshift
Surveys (MGRSs) for direct  comparison with the 2dFGRS.  The selection
criteria   for  host   and   satellite  galaxies   are  described   in
Section~\ref{sec:select}.     Section~\ref{sec:rad}    compares    the
projected,  radial distribution of  satellite galaxies  extracted from
the 2dFGRS  with those obtained  from our MGRSs, including  a detailed
discussion    of    incompleteness   effects    in    the   2dFGRS.    
Section~\ref{sec:abun} compares  the abundances of  host and satellite
galaxies  between  2dFGRS  and  MGRS.   We summarize  our  results  in
Section~\ref{sec:summ}.

\section{The Conditional Luminosity Function}
\label{sec:clf}
 
Yang, Mo \& van den Bosch (2003) and van den Bosch, Yang \& Mo (2003a)
presented a new method to link the distribution of galaxies to that of
dark  matter haloes.  This  method is  based on  modeling of  the CLF,
$\Phi(L \vert M) {\rm d}L$, which gives the average number of galaxies
with luminosity $L \pm {\rm d}L/2$ that reside in a halo of mass $M$.

The  CLF can  be constrained  by  measurements of  the abundances  and
clustering properties of  galaxies. For instance, the CLF  can be used
to compute the luminosity function of galaxies
\begin{equation}
\label{phiL}
\Phi(L) {\rm d}L = \int_0^{\infty} \Phi(L \vert M) \, n(M) \, {\rm dM}
\end{equation}
as well as the average bias of galaxies as function of luminosity
\begin{equation}
\label{biasL}
b(L) = {1 \over \Phi(L)} \int_{0}^{\infty} \Phi(L \vert M) \, 
b(M) \, n(M) \, {\rm d}M.
\end{equation}
Here $n(M)$ and  $b(M)$ are the halo mass function  and the halo bias,
respectively.

Throughout this paper we compute the halo mass function using the form
suggested by Sheth, Mo \& Tormen (2001), which has been shown to be in
excellent agreement with numerical  simulations as long as halo masses
are defined as the masses  inside a sphere with an average overdensity
of about $180$  (Jing 1998; Sheth \& Tormen  1999; Jenkins \etal 2001;
White  2002).  Therefore,  in what  follows we  consistently  use that
definition of halo mass when referring to $M$.  The halo bias, $b(M)$,
is computed using the fitting formula of Seljak \& Warren (2004).  The
linear power  spectrum of density perturbations is  computed using the
transfer function of Eisenstein  \& Hu (1998), which properly accounts
for  the  baryons,  while  the  evolved,  non-linear  power  spectrum,
required to compute the  dark matter correlation function, is computed
using the fitting formula of Smith \etal (2003).

Throughout  this paper we  assume a  flat $\Lambda$CDM  cosmology with
$\Omega_m=0.3$,  $\Omega_{\Lambda}=0.7$, $h=H_0/(100  \kmsmpc)  = 0.7$
and  with  a scale-invariant  initial  power  spectrum. Unless  stated
otherwise, we adopt a normalization of $\sigma_8=0.9$.

\subsection{Parameterization}
\label{sec:param}

The CLF is parameterized by a Schechter function:
\begin{equation}
\label{phiLM}
\Phi(L  \vert  M)  {\rm  d}L  = {\tilde{\Phi}^{*}  \over  \wLstar}  \,
\left({L \over  \wLstar}\right)^{\walpha} \, \,  {\rm exp}(-L/\wLstar)
\, {\rm d}L,
\end{equation}
where   $\wLstar   =   \wLstar(M)$,   $\walpha   =   \walpha(M)$   and
$\tilde{\Phi}^{*}  = \tilde{\Phi}^{*}(M)$  are all  functions  of halo
mass $M$. We write the average, total mass-to-light ratio of a halo of
mass $M$ as
\begin{equation}
\label{MtoLmodel}
\langle M/L \rangle_M = {1 \over 2} \,
\left({M \over L}\right)_0 \left[ \left({M \over M_1}\right)^{-\gamma_1} +
\left({M \over M_1}\right)^{\gamma_2}\right],
\end{equation}
This parameterization has four  free parameters: a characteristic mass
$M_1$, for  which the mass-to-light  ratio is equal to  $(M/L)_0$, and
two slopes,  $\gamma_1$ and $\gamma_2$,  that specify the  behavior of
$\langle M/L  \rangle$ at  the low and  high mass ends,  respectively. 
Motivated  by observations  (Bahcall,  Lubin \&  Norman 1995;  Bahcall
\etal 2000; Sanderson \& Ponman  2003), which indicate a flattening of
$\langle  M/L \rangle_M$  on  the  scale of  galaxy  clusters, we  set
$\langle  M/L \rangle_M  = (M/L)_{\rm  cl}$  for haloes  with $M  \geq
10^{14} h^{-1} \Msun$.

A  similar parameterization is used  for the characteristic luminosity
$\wLstar(M)$:
\begin{equation}
\label{LstarM}
{M \over \wLstar(M)} = {1 \over 2} \, \left({M \over L}\right)_0 \,
f(\walpha) \, \left[ \left({M \over M_1}\right)^{-\gamma_1} +
\left({M \over M_2}\right)^{\gamma_3}\right],
\end{equation}
with
\begin{equation}
\label{falpha}
f(\walpha) = {\Gamma(\walpha+2) \over \Gamma(\walpha+1,1)}.
\end{equation}
Here  $\Gamma(x)$   is  the  Gamma  function   and  $\Gamma(a,x)$  the
incomplete Gamma  function.  This parameterization  has two additional
free  parameters: a characteristic  mass $M_2$  and a  power-law slope
$\gamma_3$.   For $\walpha(M)$ we  adopt a  simple linear  function of
$\log(M)$,
\begin{equation}
\label{alphaM}
\walpha(M) = \alpha_{15} + \eta \, \log(M_{15}),
\end{equation}
with $M_{15}$ the halo mass in units of $10^{15} \msunh$, $\alpha_{15}
= \walpha(M_{15}=1)$, and $\eta$ describes the change of the faint-end
slope  $\walpha$  with  halo  mass.   Note  that  once  $\walpha$  and
$\wLstar$ are  given, the normalization $\tilde{\Phi}^{*}$  of the CLF
is  obtained through equation~(\ref{MtoLmodel}),  using the  fact that
the total (average) luminosity in a halo of mass $M$ is given by
\begin{equation}
\label{meanL}
\langle L \rangle_M = \int_{0}^{\infty}  \Phi(L \vert M) \, L \, {\rm
d}L = \tilde{\Phi}^{*} \, \wLstar \, \Gamma(\walpha+2).
\end{equation}
Finally, we introduce the mass  scale $M_{\rm min}$ below which we set
the CLF to zero; i.e., we assume that no stars form inside haloes with
$M < M_{\rm min}$.  Motivated by reionization considerations (see Yang
\etal 2003 for  details) we adopt $M_{\rm min}  = 10^{9} h^{-1} \Msun$
throughout.

\subsection{Parameter Fitting}
\label{sec:mcm}

The CLF,  as specified above,  has a total  of 8 free  parameters: two
characteristic masses; $M_1$ and $M_2$, three parameters that describe
the various  mass-dependencies $\gamma_1$, $\gamma_3$  and $\eta$, two
normalization for the mass-to-light  ratio, $(M/L)_0$ and $\langle M/L
\rangle_{\rm  cl}$,  and  a  normalization  of  the  faint-end  slope,
$\alpha_{15}$.  Note  that $\gamma_2$  is not a  free parameter  as it
derives from requiring continuity in $\langle M/L \rangle_M$ across $M
= 10^{14}  h^{-1} \Msun$. The  data that we  use to constrain  the CLF
consists of  the 2dFGRS luminosity  function of Madgwick  \etal (2002)
and the  galaxy-galaxy correlation  lengths as function  of luminosity
obtained   from   the   2dFGRS   by  Norberg   \etal   (2002a).    The
goodness-of-fit  of each  model  is described  by  its $\chi^2$  value
defined by $\chi^2 = \chi^2_{\Phi} + \chi^2_{r_0}$ with
\begin{equation}
\label{chisqLF}
\chi^2_{\Phi} = \sum_{i=1}^{N_{\Phi}}
\left[ {\Phi(L_i) - \hat{\Phi}(L_i) \over \Delta \hat{\Phi}(L_i)} \right]^2,
\end{equation}
and
\begin{equation}
\label{chisqr0}
\chi^2_{r_0} = \sum_{i=1}^{N_{r}}
\left[ {r_0(L_i) - \hat{r}_0(L_i) \over \Delta \hat{r}_0(L_i)} \right]^2,
\end{equation}
Here  $\hat{\Phi}$ and  $\hat{r}_0$ are  the observed  quantities, and
$N_{\Phi}=35$ and $N_{r}=8$  are the number of data  points for the LF
and the correlation lengths, respectively. 

In Yang  \etal (2003) and van  den Bosch \etal (2003a)  we presented a
number of CLFs  that accurately fit these data and  that were based on
different assumptions  regarding the free parameters.   Motivated by a
number of independent observational constraints, the majority of these
models were constrained to have  $(M/L)_{\rm cl} \simeq 500 h \MLsun$. 
However, subsequent  studies have shown that these  CLF models predict
too many rich galaxy groups  (Yang \etal 2004b), and pairwise peculiar
velocity dispersions that are too high (Yang \etal 2004a).  Both these
problems   are  alleviated   by   adopting  a   much  higher   cluster
mass-to-light ratio of $(M/L)_{\rm cl} \simeq 900 h \MLsun$.

In  this paper  we adopt  a  different approach.   Rather than  fixing
$(M/L)_{\rm cl}$ at a preferred value and using a minimization routine
to  search  our multi-dimensional  parameter  space  for the  best-fit
parameters,  we  follow  Yan,  Madgwick  \& White  (2003)  and  use  a
Monte-Carlo  Markov  Chain  (hereafter  MCMC) to  fully  describe  the
likelihood  function in  our multi-dimensional  parameter  space. This
will allow  us to more  accurately investigate the freedom  in cluster
mass-to-light  ratio. Readers  not  familiar with,  or interested  in,
Monte-Carlo Markov Chains are referred to Gamerman (1997) for details.
\begin{figure*}
\centerline{\psfig{figure=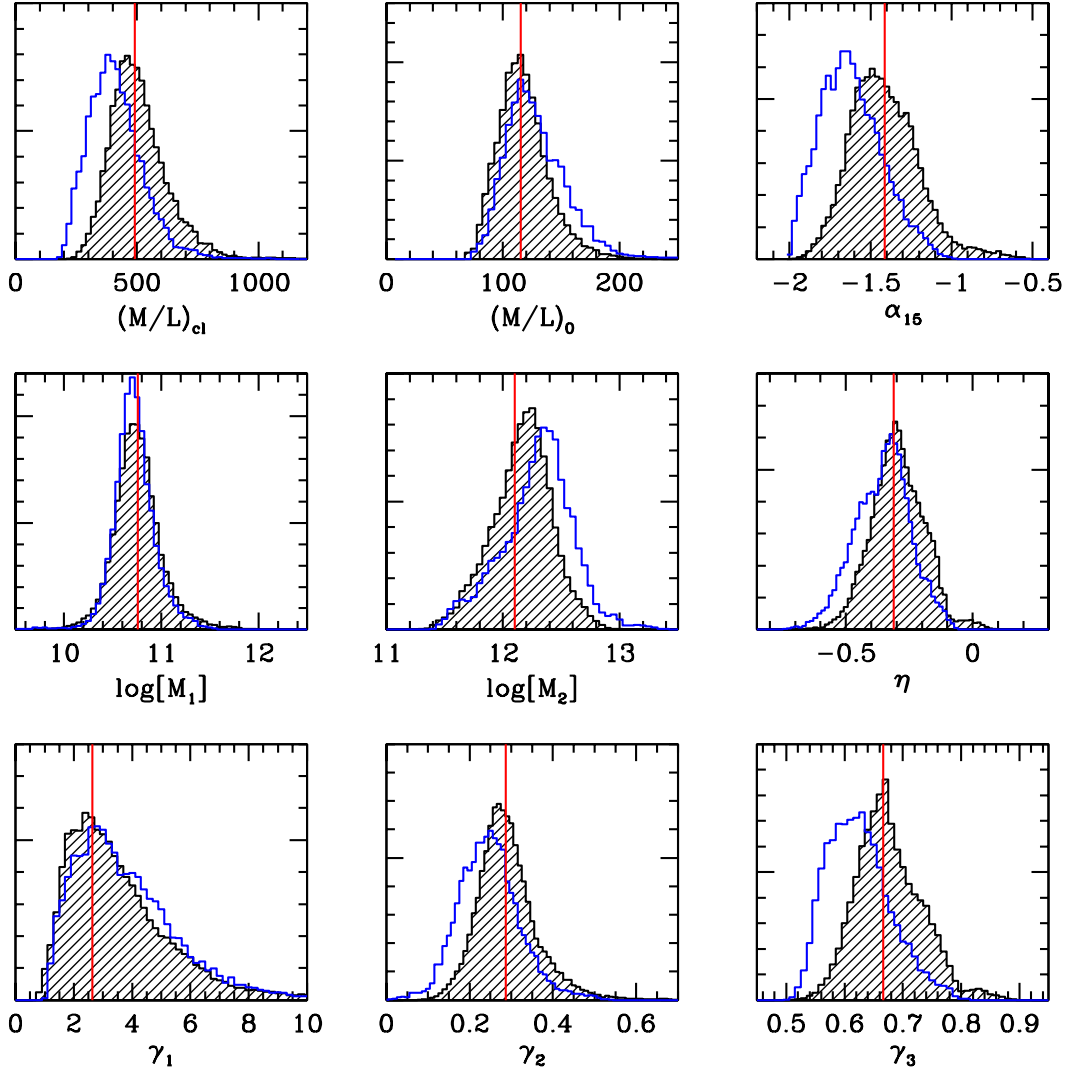,width=0.8\hdsize}}
\caption{Constraints on the nine CLF parameters obtained from Monte-Carlo
  Markov   Chains  with   2000  independent   samples.    The  hatched
  (non-hatched)  histograms  correspond  to  $\Lambda$CDM  concordance
  cosmologies  with  $\sigma_8=0.9$  ($0.7$), respectively.   Vertical
  lines  indicate  the  best-fit  parameters  for  the  $\sigma_8=0.9$
  cosmology. The  median and 68 percent confidence  intervals of these
  distributions are listed in Table~1. Masses and mass-to-light ratios
  are in units of $h^{-1} \Msun$ and $h \MLsun$, respectively.}
\label{fig:histo}
\end{figure*}

We start  our MCMC  from model~D  in van den  Bosch \etal  (2004a) and
allow  a  `burn-in' of  10.000  random walk  steps  for  the chain  to
equilibrate  in the likelihood  space. At  any point  in the  chain we
generate a  new trial model  by drawing the  shifts in its  eight free
parameter   from  eight   independent   Gaussian  distributions.   The
probability of accepting the trial model is
\begin{equation}
\label{probaccept}
P_{\rm accept} = \left\{ \begin{array}{ll}
1.0 & \mbox{if $\chi^2_{\rm new} < \chi^2_{\rm old}$} \\
{\rm exp}[-(\chi^2_{\rm new}-\chi^2_{\rm old})/2] & \mbox{if 
$\chi^2_{\rm new} \geq \chi^2_{\rm old}$} \end{array} \right.
\end{equation}
with    the   $\chi^2$    measures   given    by   eq.~(\ref{chisqLF})
and~(\ref{chisqr0}).
\begin{figure*}
\centerline{\psfig{figure=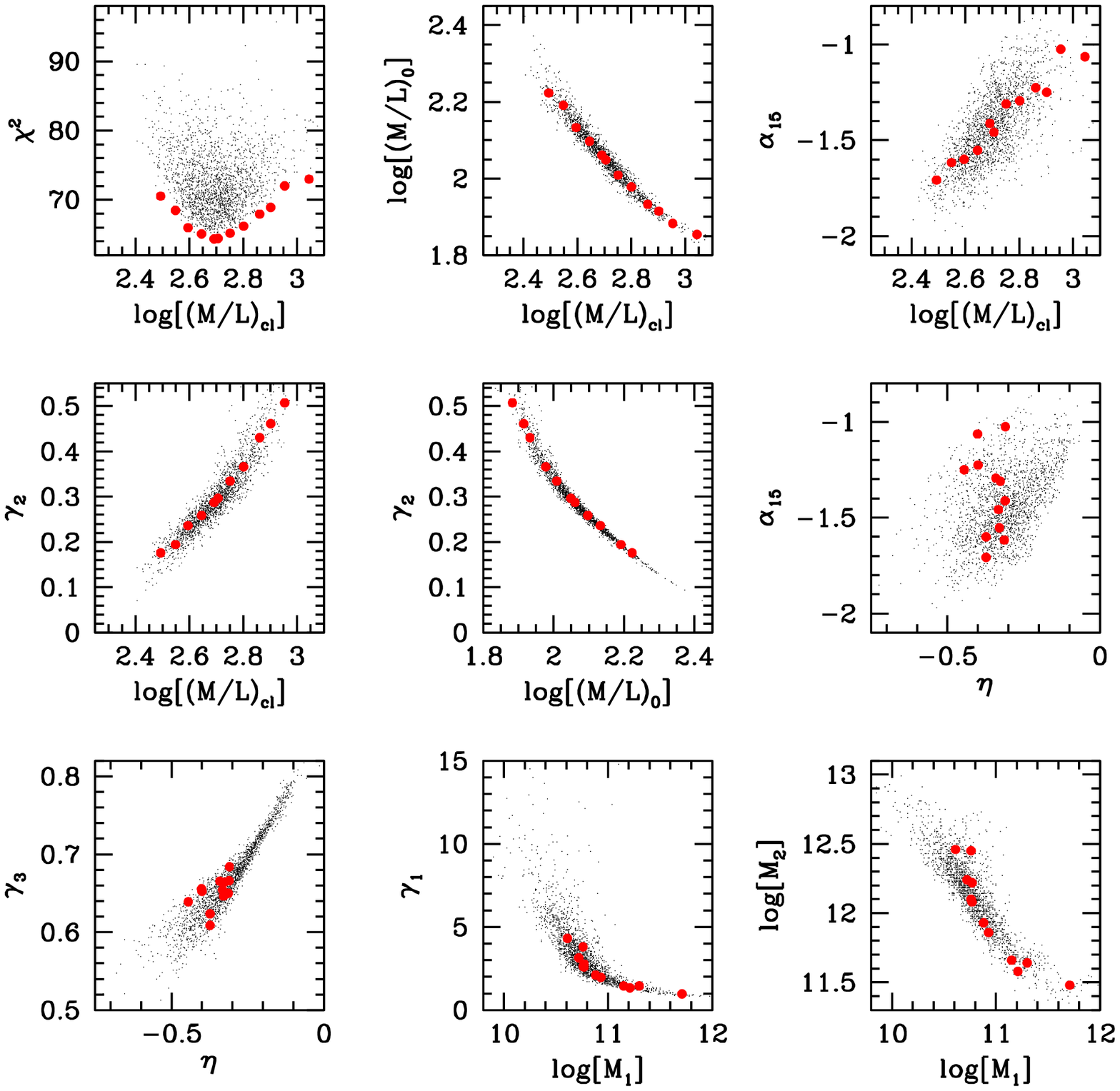,width=0.9\hdsize}}
\caption{Correlations between various CLF parameters of the 2000 samples
  in  the $\sigma_8=0.9$  MCMC  (thin dots).   The  thick, solid  dots
  correspond  to  the  twelve  CLF  models  listed  in  Table~1  which
  represent  the  best-fit  models  in  twelve  intervals  of  cluster
  mass-to-light  ratio, $(M/L)_{\rm  cl}$.   Masses and  mass-to-light
  ratios are in units of $h^{-1} \Msun$ and $h \MLsun$, respectively.}
\label{fig:scatter}
\end{figure*}

We construct a MCMC of  $40$ million steps, with an average acceptance
rate of $\sim 12$ percent.  In order to suppress the correlation power
between neighboring models in the chain, we thin the chain by a factor
$20.000$.   This  results  in   a  final  MCMC  consisting  of  $2000$
independent   models   that  properly   sample   the  full   posterior
distribution.  The hatched histograms in Fig.~\ref{fig:histo} plot the
resulting  distributions  of  parameters,  with  the  best-fit  values
indicated by  a vertical  line. The median  and 68  percent confidence
intervals  of  these  distributions   are  listed  in  Table~1  (model
$\Lambda_{0.9}$).  Note that these  parameters are similar to those of
models A--D  presented in  van den Bosch  \etal (2004a)\footnote{Small
  differences  are due  to the  fact that  here we  use the  halo bias
  fitting function  of Seljak  \& Warren (2004),  rather than  that of
  Sheth, Mo \& Tormen (2001),  and the transfer function of Eisenstein
  \&  Hu  (1998),  rather  than  that of  Efstathiou,  Bond  \&  White
  (1992).}.
\begin{figure*}
\centerline{\psfig{figure=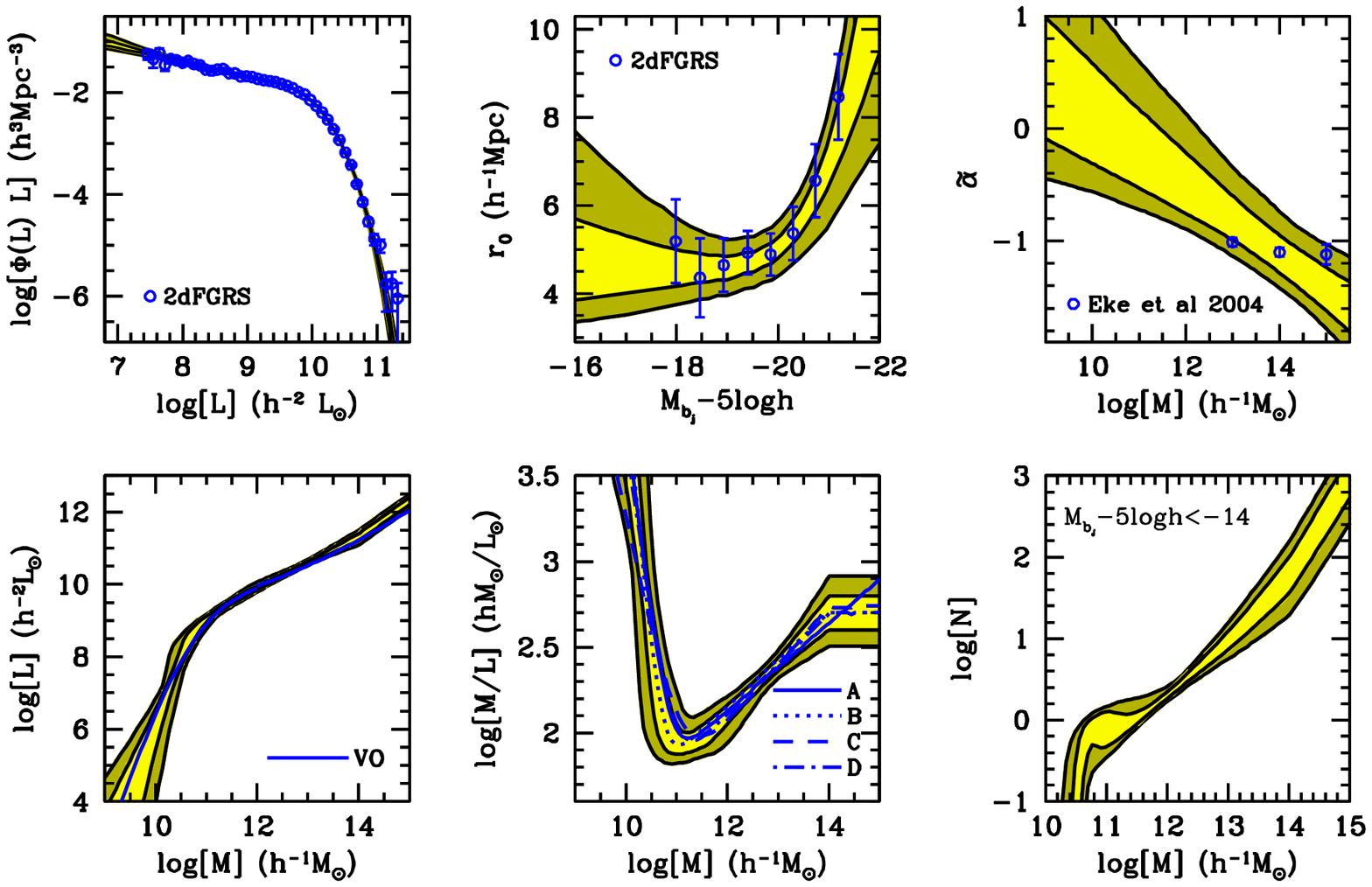,width=\hdsize}}
\caption{Posterior constraints on a number of quantities computed from
  the  $\sigma_8=0.9$  MCMC.  The  contours  show  the  68\% and  99\%
  confidence  limits from the  marginalized distribution.   {\it Upper
    left-hand  panel:} The  galaxy luminosity  function;  open circles
  with  errorbars correspond to  the 2dFGRS  data from  Madgwick \etal
  (2002).  {\it Upper middle panel:} galaxy-galaxy correlation lengths
  as  function  of absolute  magnitude;  open  circles with  errorbars
  correspond  to the  2dFGRS data  from Norberg  \etal  (2002a).  {\it
    Upper  right-hand  panel:}  the  faint-end  slope of  the  CLF  as
  function of halo mass; open circles with errorbars correspond to the
  2dFGRS data from Eke \etal  (2004). {\it Lower left-hand panel:} the
  total luminosity per  halo as function of halo  mass. The solid blue
  line corresponds  to the  model of Vale  \& Ostriker (2004),  and is
  shown  for  comparison.   {\it  Lower  middle  panel:}  the  average
  mass-to-light ratio as function of halo mass. The various blue lines
  correspond  to  models A--D  of  van  den  Bosch \etal  (2003a),  as
  indicated.   {\it Lower  right-hand  panel:} The  average number  of
  galaxies with $M_{b_J}  - 5 {\rm log} h < -14$  per halo as function
  of halo mass. See text for a detailed discussion.}
\label{fig:mcmccl}
\end{figure*}

In Fig.~\ref{fig:scatter} we  show scatter plots for some  of the most
tightly correlated  pairs of parameters. The number  density of points
in these  plots is directly  proportional to the probability  density. 
The solid circles correspond to twelve specific models to be discussed
in more  detail below.  The  most pronounced correlations  are between
$(M/L)_{\rm  cl}$, $(M/L)_0$  and $\gamma_2$:  increasing  the cluster
mass-to-light ratio results in a decrease of the minimum mass-to-light
ratio $(M/L)_0$.  This simply  reflects the conservation of the number
of  galaxies  (constrained  by  the  LF): putting  fewer  galaxies  in
clusters requires an  increase in the occupation number  in lower mass
haloes.   Since   $\gamma_2$  reflects  the  slope   of  $\langle  M/L
\rangle_M$ in  between $M_1$ and the  cluster scale, it  is clear that
$\gamma_2$  has to  increase with  increasing $(M/L)_{\rm  cl}$ and/or
decreasing $(M/L)_0$.  Another  important correlation is between $M_1$
and $\gamma_1$, which is a direct consequence of the shallow faint-end
slope of  the galaxy LF. Because  of the much steeper  low-mass end of
the halo mass  function, $\langle M/L \rangle_M$ has  to increase with
decreasing mass.   Lowering $M_1$ results  in a steeper  increase, and
therefore a larger value of $\gamma_1$.

Fig.~\ref{fig:mcmccl}  shows confidence  levels on  various quantities
computed from  our MCMC. The blue  open circles with  errorbars in the
upper left and upper middle panels indicate the data used to constrain
the  models.  The  colored  areas  indicate  the  68  and  99  percent
confidence levels  on $\Phi(L)$ and  $r_0(L)$ computed from the  MCMC. 
Note the  good agreement  with the data,  indicating that the  CLF can
accurately match the observed  abundances and clustering properties of
galaxies  in the  2dFGRS.  We  emphasize that  this is  not  a trivial
result, as  the data can only  be fitted for a  certain combination of
cosmological parameters (van den Bosch \etal 2003b).

The  upper   right-hand  panel  of   Fig.~\ref{fig:mcmccl}  plots  the
faint-end slope of the CLF as function of halo mass.  The data clearly
favors models in which $\walpha$  increases with decreasing halo mass. 
At around  the cluster scale  the models favor fairly  steep faint-end
slopes  with $-1.0  \lta \walpha  \lta  -1.5$, which  is in  excellent
agreement with  independent studies of  the luminosity functions  in a
number  of individual  clusters  (e.g., Sandage,  Bingelli \&  Tammann
1985; Beijersbergen \etal 2002;  Trentham \& Hodgkin 2002; Trentham \&
Tully  2002).    The  blue   open  circles  with   errorbars  indicate
$\walpha(M)$ for three different halo masses $M$ obtained by Eke \etal
(2004) from  an analysis  of groups  in the  2dFGRS, and  are  in good
agreement with our model.

The lower left-hand panel  of Fig.~\ref{fig:mcmccl} plots the relation
between  halo  mass  $M$  and  the  total  halo  luminosity  $L$,  the
expectation  value  of  which   follows  from  the  CLF  according  to
eq.~(\ref{meanL}).   Note  that the  confidence  levels are  extremely
tight, especially  for the more  massive haloes.  The  $L(M)$ relation
reveals a  dramatic break at around  $M = M_1 \simeq  7 \times 10^{10}
h^{-1} \Msun$, with $L \propto M^{3.5  \pm 1.5}$ for $M << M_1$ and $L
\propto M^{0.73 \pm 0.08}$ for $M >> M_1$ (errorbars are obtained from
the confidence levels on  $\gamma_1$ and $\gamma_2$, respectively). In
a recent  study, Vale  \& Ostriker (2004),  under the assumption  of a
monotonic $L(M)$, obtained the relation  between light and mass from a
comparison of the galaxy luminosity function with the {\it total} halo
mass function  (counting both parent  and subhaloes).  They  find that
$L(M)$ changes from $L \propto  M^{4}$ to $L \propto M^{0.9}$, in good
agreement  with our  results.   This  is also  evident  from a  direct
comparison of their  results (blue solid line in  the lower left panel
of  Fig.~\ref{fig:mcmccl})  with  ours,  which  shows  almost  perfect
agreement. It  is extremely reassuring that two  such wildly different
methods yield results  in such good agreement. This  combined with the
extremely  tight  confidence levels  obtained  from  our CLF  analysis
suggests  that  we have  established  a  remarkably robust  connection
between  galaxy light  and halo  mass  (at least  for the  concordance
cosmology adopted here).

The   lower,   middle  panel   of   Fig.~\ref{fig:mcmccl}  plots   the
corresponding  mass-to-light ratios  as  function of  halo mass.   The
pronounced minimum  in $M/L$ indicates  that galaxy formation  is most
efficient in haloes with masses  in the range $5 \times 10^{10} h^{-1}
\Msun \lta M \lta 10^{12} h^{-1} \Msun$.  For less massive haloes, the
mass-to-light ratio  increases drastically with  decreasing halo mass,
which is required  in order to bring the steep slope  of the halo mass
function at low $M$ in agreement with the relatively shallow faint-end
slope of the observed LF.  It indicates that galaxy formation needs to
become extremely inefficient  in haloes with $M \lta  5 \times 10^{10}
h^{-1} \Msun$ in order to  prevent an overabundance of faint galaxies. 
The increase  in $\langle M/L  \rangle_M$ from $M \sim  10^{11} h^{-1}
\Msun$  to  $M \sim  10^{14}  h^{-1}  \Msun$  is associated  with  the
decreasing ability of the gas to cool with increasing halo mass (e.g.,
White  \& Rees  1978).  Finally,  the  sharp turn-over  to a  constant
$\langle M/L \rangle_M$ for haloes  with $M \geq 10^{14} h^{-1} \Msun$
is   a   direct   reflection   of  our   CLF   parameterization   (see
Section~\ref{sec:param}).  The various lines correspond to models A--D
from van den Bosch \etal  (2003a), and are in excellent agreement with
the confidence  intervals obtained here.  This indicates  that the CLF
models used  in our previous work  (Yang \etal 2004a,b;  van den Bosch
\etal 2003a,b; van den Bosch  \etal 2004a,b; Mo \etal 2004; Wang \etal
2004) are perfectly consistent with the parameter constraints obtained
here using the MCMC.

Finally,  the lower  right-hand panel  of  Fig.~\ref{fig:mcmccl} plots
predictions for the number of galaxies  (with $M_{b_J} - 5 {\rm log} h
< -14$) as function of halo mass. These derive from the CLF according
to
\begin{equation}
\label{totN}
\langle N \rangle_M = \int_{L_{\rm min}}^{\infty} \Phi(L \vert M) \, {\rm d}L
\end{equation}
with $L_{\rm  min}$ the minimum luminosity considered  ($L_{\rm min} =
5.2 \times 10^7 h^{-2} \Lsun$  in our case).  Although the exact shape
and normalization  of $\langle N  \rangle_M$ depends on  $L_{\rm min}$
(e.g.,  van den Bosch  \etal 2003a),  the power-law  behavior combined
with a  shoulder plus  break at low  $M$ is  in good agreement  with a
number  of studies  based  on halo  occupation  numbers (Seljak  2000;
Scranton  2003; Berlind  \etal 2003;  Magliocchetti \&  Porciani 2003;
Kravtsov \etal 2003).

\section{Mock Galaxy Redshift Surveys}
\label{sec:mock}

In order  to properly interpret the  2dFGRS data on  the abundance and
radial distribution  of satellite galaxies we  construct detailed mock
galaxy redshift  surveys (hereafter  MGRS).  These have  the advantage
that (i) we know exactly the  {\it true} abundances and the {\it true}
radial  distributions,  (ii)  we  can  model the  effects  of  various
observational  biases,   and  (iii)  we  can  use   exactly  the  same
host/satellite  selection  criteria  as  for the  2dFGRS,  making  the
comparison with the data straightforward.

To  construct MGRSs two  ingredients are  required; a  distribution of
dark  matter haloes  and a  description of  how galaxies  of different
luminosity occupy  haloes of  different mass.  For  the former  we use
large  numerical  simulations,  and  for the  latter  the  conditional
luminosity function. Ideally we would construct a MGRS for each of the
2000 models in our MCMC.   Since this is computationally too expensive
we adopt an  alternative method. We determine the  minimum and maximum
values of $(M/L)_{\rm cl}$ in  the entire MCMC, and split the interval
in twelve equal size logarithmic bins.  For each of these bins we then
determine the model that yields the lowest $\chi^2$. The parameters of
the resulting twelve  models are listed in Table~1,  and are indicated
by  thick solid  dots  in Fig.~\ref{fig:scatter}.  For  each of  these
twelve  CLFs  we  construct  a  MGRS,  which  we  use  to  assess  the
uncertainties on  the abundances and radial  distribution of satellite
galaxies due to the uncertainties in $(M/L)_{\rm cl}$. In what follows
we refer to MGRS based on CLF  $n$ as M$n$, where $n$ is the ID listed
in Column~(1) of Table~1.
\begin{table*}
\caption{Conditional luminosity function parameters.}
\begin{tabular}{ccccccccccc}
   \hline
ID & $(M/L)_{cl}$ & $(M/L)_0$ & log$M_1$ & log$M_2$ &
$\gamma_1$ & $\gamma_2$ & $\gamma_3$ & $\alpha_{15}$ & $\eta$ & 
$\chi^2$ \\
 (1) & (2) & (3) & (4) & (5) & (6) & (7) & 
(8) & (9) & (10) & (11) \\
\hline\hline
 1 &  $311.2$ & $167.1$ & $10.76$ & $12.45$ & $3.803$ & $0.176$ & $0.609$ & $-1.71$ & $-0.374$ & $70.53$ \\
 2 &  $353.2$ & $155.2$ & $10.61$ & $12.46$ & $4.310$ & $0.194$ & $0.650$ & $-1.62$ & $-0.314$ & $68.46$ \\
 3 &  $393.6$ & $135.8$ & $10.77$ & $12.22$ & $2.770$ & $0.236$ & $0.624$ & $-1.60$ & $-0.373$ & $65.97$ \\
 4 &  $441.6$ & $125.0$ & $10.72$ & $12.24$ & $3.141$ & $0.259$ & $0.646$ & $-1.55$ & $-0.330$ & $65.09$ \\
 5 &  $490.9$ & $115.1$ & $10.76$ & $12.10$ & $2.639$ & $0.287$ & $0.666$ & $-1.41$ & $-0.312$ & $64.34$ \\
 6 &  $508.2$ & $111.9$ & $10.77$ & $12.08$ & $2.594$ & $0.297$ & $0.654$ & $-1.46$ & $-0.333$ & $64.42$ \\
 7 &  $563.6$ & $102.1$ & $10.88$ & $11.93$ & $2.106$ & $0.334$ & $0.664$ & $-1.31$ & $-0.327$ & $65.18$ \\
 8 &  $632.4$ &  $95.1$ & $10.93$ & $11.86$ & $1.970$ & $0.366$ & $0.665$ & $-1.29$ & $-0.342$ & $66.20$ \\
 9 &  $726.1$ &  $85.7$ & $11.15$ & $11.66$ & $1.451$ & $0.431$ & $0.653$ & $-1.23$ & $-0.400$ & $67.95$ \\
10 &  $798.0$ &  $82.2$ & $11.21$ & $11.58$ & $1.331$ & $0.461$ & $0.639$ & $-1.25$ & $-0.445$ & $68.91$ \\
11 &  $899.5$ &  $76.4$ & $11.30$ & $11.64$ & $1.444$ & $0.507$ & $0.684$ & $-1.10$ & $-0.310$ & $72.01$ \\
12 & $1106.6$ &  $71.4$ & $11.71$ & $11.48$ & $0.976$ & $0.652$ & $0.656$ & $-1.11$ & $-0.401$ & $72.99$ \\
\hline
 $\Lambda_{0.9}$ &  $500^{+130}_{-100}$ & $115^{+23}_{-20}$ & $10.76^{+0.25}_{-0.21}$ & $12.15^{+0.24}_{-0.31}$ & $2.92^{+2.11}_{-1.15}$ & $0.29^{+0.08}_{-0.06}$ & $0.66^{+0.06}_{-0.04}$ & $-1.44^{+0.21}_{-0.18}$ & $-0.31^{+0.10}_{-0.10}$ & -- \\
\hline
 $\Lambda_{0.7}$ &  $400^{+120}_{-100}$ & $125^{+30}_{-21}$ & $10.71^{+0.21}_{-0.17}$ & $12.32^{+0.26}_{-0.35}$ & $3.46^{+2.18}_{-1.35}$ & $0.24^{+0.08}_{-0.07}$ & $0.62^{+0.06}_{-0.05}$ & $-1.64^{+0.21}_{-0.17}$ & $-0.35^{+0.10}_{-0.12}$ & -- \\
\hline
 $\Lambda_{0.7}^{\rm ab}$ &  $570^{+40}_{-40}$   &  $98^{+3}_{-4}$   & $10.86^{+0.04}_{-0.04}$ & $11.96^{+0.10}_{-0.11}$ & $2.45^{+0.55}_{-0.37}$ & $0.34^{+0.02}_{-0.01}$ & $0.63^{+0.03}_{-0.05}$ & $-1.35^{+0.04}_{-0.25}$ & $-0.35^{+0.08}_{-0.11}$ & -- \\
\hline
\end{tabular}
\medskip

\begin{minipage}{\hdsize}
  Parameters of CLF models.  Column~(1) lists the ID by which we refer
  to  each  CLF in  the  text.   Columns~(2)  to~(10) list  the  model
  parameters, and  column~(11) the value of $\chi^2  = \chi^2_{\Phi} +
  \chi^2_{r_0}$.   Masses  and  mass-to-light  ratios are  in  $h^{-1}
  \Msun$  and $h \MLsun$,  respectively. The  first twelve  lines (IDs
  1--12)  correspond  to  the   best-fit  models  extracted  from  the
  $\sigma_8=0.9$ MCMC  for different  bins in $(M/L)_{\rm  cl}$.  They
  are shown  as thick solid dots in  Fig.~\ref{fig:scatter}.  The last
  three lines list the median  and 68 percent confidence levels of the
  parameter probability distributions obtained from the MCMCs.  Models
  $\Lambda_{0.9}$ and  $\Lambda_{0.7}$ correspond  to the MCMC  of the
  $\Lambda$CDM  cosmologies  with  $\sigma_8=0.9$ and  $\sigma_8=0.7$,
  respectively. Model  $\Lambda_{0.7}^{\rm ab}$  is the same  as Model
  $\Lambda_{0.7}$, except  that we have  weighted the MCMC  samples by
  ${\rm exp}(-\chi^2_{\rm ab}/2)$ (see Section~\ref{sec:sigma}).
\end{minipage}

\end{table*}

Details about  the construction of the  MGRSs can be found  in van den
Bosch \etal (2004b) and Yang  \etal (2004a). Here we briefly summarize
the  main ingredients. The  MGRSs are  constructed by  populating dark
matter  haloes  in a  stack  of  large  numerical simulations,  kindly
provided to us  by Y.P.  Jing (see Jing \& Suto  2002).  The number of
galaxies per halo, and their  luminosities, are obtained from the CLF. 
We assume  that the brightest galaxy  in each halo resides  at rest at
the  halo  center,  while  all  other galaxies  are  in  an  isotropic
steady-state   equilibrium  and   follow  a   radial   number  density
distribution given by
\begin{equation}
\label{nsatr}
n_{\rm sat}(r) \propto \left( {r \over {\cal R} r_s} \right)^{-\alpha}
\left( 1 + {r \over {\cal R} r_s} \right)^{\alpha-3}
\end{equation}
(limited to $r \leq r_{\rm vir}$) with $r_s$ the characteristic radius
of the NFW  profile, and $\alpha$ and ${\cal R}$  two free parameters. 
Unless specifically stated otherwise we  adopt $\alpha = {\cal R} = 1$
for  which  the  number  density distribution  of  satellite  galaxies
exactly follows the dark matter mass distribution.
\begin{figure}
\centerline{\psfig{figure=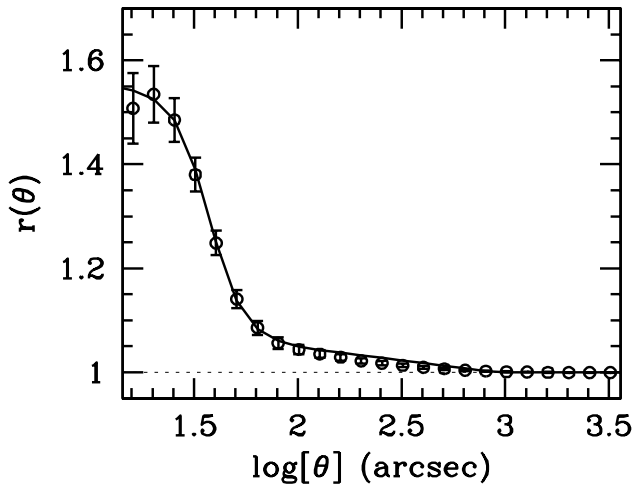,width=\hssize}}
\caption{The solid line corresponds to the $r(\theta)$ (see text for
  definition) obtained  from the 2dFGRS by Hawkins  \etal (2003). Note
  that $r$ is significantly larger than unity for $\theta \lta 100''$,
  indicating  that the  2dFGRS has  missed a  significant  fraction of
  close  pairs (largely due  to fiber  collisions).  The  open circles
  with errorbars  correspond to the $r(\theta)$ obtained  for MGRS M6,
  after  we have  removed close  pairs.  The  good agreement  with the
  results from  the 2dFGRS indicates  that we have constructed  a MGRS
  with the  same deficiency of close  pairs as in the  data. Note that
  this does  not correct for  a possible close-pair  incompleteness in
  the    {\it     parent}    catalogue    of     the    2dFGRS    (see
  Section~\ref{sec:project}).}
\label{fig:rtheta}
\end{figure}

As  described in  detail in  Colless  \etal (2001)  and Norberg  \etal
(2002b), the {\it parent} catalogue  of the 2dFGRS, which is extracted
from the APM catalogue, is only 91\% complete. In addition, because of
problems with  the tiling  strategy and because  of noisy  spectra the
level  of completeness  of the  2dFGRS depends  on  position, apparent
magnitude,  and pair-separation.   In  the MGRSs  used  in Yang  \etal
(2004a),  Wang  \etal  (2004)  and  van den  Bosch  \etal  (2004b)  we
corrected  for  the  average   incompleteness  of  the  2dFGRS  parent
catalogue and for the position and magnitude dependent incompleteness,
using detailed completeness  maps provided by the 2dFGRS  team, but no
correction  was  made for  the  pair-separation  dependency.  For  the
purpose  of   investigating  the  radial   distribution  of  satellite
galaxies, however, this close pair incompleteness needs to be properly
accounted  for.  Hawkins  \etal (2003)  computed the  ratio $r(\theta)
\equiv  [1 +  w_p(\theta)] /  [1 +  w_z(\theta)]$ between  the angular
correlation functions  of the 2dFGRS  parent catalogue, $w_p(\theta)$,
and that of the final  redshift survey, $w_z(\theta)$.  They found $r$
to be  significantly larger than unity, corresponding  to a deficiency
of  pairs, for  $\theta  \lta 100''$.   In  an attempt  to model  this
close-pair  deficiency in  our  MGRSs we  proceed  as follows.   After
having   corrected   for   the   position-   and   magnitude-dependent
incompleteness,  we compute the  angular separations  $\theta$ between
all  galaxy   pairs  and  remove  galaxies  based   on  a  probability
$p(\theta)$, which we  tune (by trial and error)  so that we reproduce
the $r(\theta)$ obtained  by Hawkins \etal (2003).  As  a last step we
then  remove a  number of  galaxies  completely at  random such  that,
together with the galaxies removed because of their angular separation
to neighbors,  we have removed 9  percent of all  mock galaxies.  This
mimics the incompleteness  level of the parent catalogue\footnote{Note
  that  the  sequential  order   in  which  the  various  completeness
  corrections  have been  applied is  not entirely  correct.  However,
  numerous  tests have shown  that this  has an  absolutely negligible
  effect on  our results.}.   The solid line  in Fig.~\ref{fig:rtheta}
shows $r(\theta)$ for the 2dFGRS  as obtained by Hawkins \etal (2003),
while  open  circles  with  errorbars correspond  to  the  $r(\theta)$
obtained  from a comparison  of the  angular correlation  functions in
MGRS  M6  before and  after  the removal  of  close  pairs.  The  good
agreement with the 2dFGRS results demonstrates that we have managed to
construct MGRSs with the same deficiency of close pairs as in the real
2dFGRS. All MGRSs discussed in this paper have been corrected for this
close-pair incompleteness.

\section{Selecting Host and Satellite Galaxies}
\label{sec:select}

A galaxy is considered a potential host galaxy if it is at least $f_h$
times brighter than any other galaxy within a volume specified by $R_p
< R_h$ and  $\vert \Delta V \vert < (\Delta V)_h$.   Here $R_p$ is the
separation projected on the sky at the distance of the candidate host,
and $\Delta V$ is  the line-of-sight velocity difference.  Around each
potential  host  galaxy,  satellite  galaxies  are  defined  as  those
galaxies that  are at  least $f_s$ times  fainter than their  host and
located within a  volume with $R_p < R_s$ and $\vert  \Delta V \vert <
(\Delta V)_s$.  Host galaxies with zero satellite galaxies are removed
from the list of hosts.

In total,  the selection of hosts  and satellites thus  depends on six
free  parameters:  $R_h$, $(\Delta  V)_h$  and  $f_h$  to specify  the
population of  host galaxies, and  $R_s$, $(\Delta V)_s$ and  $f_s$ to
specify the  satellite galaxies.  These parameters  also determine the
number of  interlopers (defined as a galaxy  not physically associated
with the halo of the host  galaxy) and non-central hosts (defined as a
host  galaxy that  is not  the brightest,  central galaxy  in  its own
halo).   Minimizing the  number of  interlopers  requires sufficiently
small $R_s$  and $(\Delta V)_s$. Minimizing the  number of non-central
hosts  requires  one  to   choose  $R_h$,  $(\Delta  V)_h$  and  $f_h$
sufficiently   large.    Of  course,   each   of  these   restrictions
dramatically reduces  the number of both hosts  and satellites, making
the statistical estimates more and more noisy.

In van  den Bosch  \etal (2004b)  we used our  MGRS to  optimize these
selection criteria,  aiming for large numbers of  hosts and satellites
(to increase signal-to-noise), a  small fraction of interlopers, and a
small fraction  of non-central  hosts. It was  shown that  an adaptive
selection  criterion, for which  $R_h$, $(\Delta  V)_h$ and  $R_s$ are
made  dependent of  the  luminosity of  the  host-candidate, was  most
successful.  Motivated  by these  findings we adopt:  $f_h = f_s  = 1$,
$(\Delta V)_h =  1000 \sigma_{200} \kms$, $(\Delta V)_s  = 2000 \kms$,
$R_h =  0.8 \sigma_{200}  h^{-1} \Mpc$, and  $R_s =  0.15 \sigma_{200}
h^{-1}   \Mpc$.   Here  $\sigma_{200}$   is  the   satellite  velocity
dispersion, $\sigma_{\rm sat}(L_{\rm host})$,  in units of $200 \kms$,
for which we adopt
\begin{equation}
\label{siglum}
{\rm log} \sigma_{\rm sat} =  {\rm log} \sigma_{10} + a_1
{\rm log} L_{10}  + a_2 ({\rm log} L_{10})^2 
\end{equation}
Here $L_{10} = L_{\rm host} / 10^{10} h^{-2} \Lsun$ and $\sigma_{10} =
\sigma_{\rm  sat}(L_{10})$. We set  $\sigma_{10} =  200 \kms$,  $a_1 =
0.5$ and  $a_2 = 0.1$, which  is close to the  parameters obtained for
the 2dFGRS by van den Bosch \etal (2004b). Note that details regarding
the selection  criteria of  host and satellite  galaxies are  not very
important as long as the data and model (MGRS) are treated in the same
way.  

For the  data we use the  final, public data release  from the 2dFGRS,
restricting  ourselves only to  galaxies with  redshifts $0.01  \leq z
\leq  0.15$  in  the  North  Galactic Pole  and  South  Galactic  Pole
subsamples with a redshift quality  parameter $q \geq 3$.  This leaves
a grand total  of $146735$ galaxies with a  typical rms redshift error
of $85  \kms$ (Colless \etal 2001).  Absolute  magnitudes for galaxies
in the 2dFGRS  are computed using the K-corrections  of Madgwick \etal
(2002).  Applying our host-satellite  selection criterion  yields 8737
host galaxies and 13738 satellite galaxies.
\begin{figure*}
\centerline{\psfig{figure=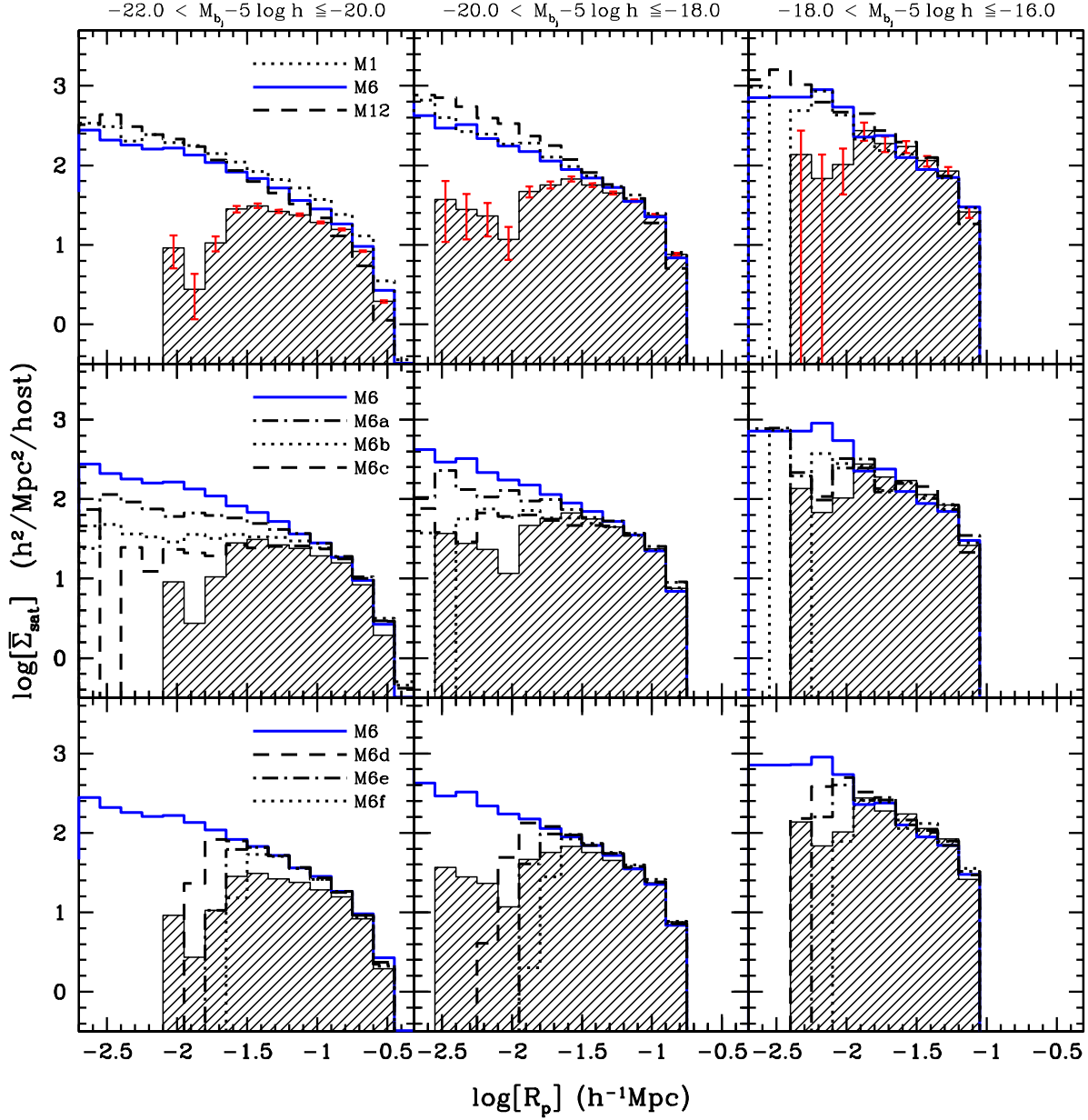,width=0.9\hdsize}}
\caption{The projected number density distributions of satellite
  galaxies  as function  of projected  radius. Results  are  shown for
  three different bins in host-galaxy luminosity, indicated at the top
  of each column. In each  panel, the hatched histogram corresponds to
  the 2dFGRS data, while the various lines have been obtained from the
  MGRSs. The  errorbars (only shown  in the upper panels  for clarity)
  indicate  Poisson  errors  in  the  numbers of  host  and  satellite
  galaxies. The  upper panels compare  the 2dFGRS data to  three MGRSs
  (M1,  M6,   and  M12)  that   span  the  entire  range   in  cluster
  mass-to-light ratio (see Table~1  for parameters). The middle row of
  panels compare the 2dFGRS data to MGRSs based on CLF 6, but in which
  the intrinsic, spatial distribution of satellite galaxies is modeled
  with a constant number  density core (eq.~[\ref{nsatr}] with $\alpha
  =  0$) with  ${\cal  R}  = 1$  (M6a),  2 (M6b),  and  3 (M6c).   For
  comparison the fiducial  MGRS M6, for which $\alpha =  {\cal R} = 1$
  is also shown (solid lines).   Finally, the lower panels compare the
  2dFGRS data to MGRSs based on  CLF 6, but in which we have corrected
  for  the size-projection  incompleteness using  $R_{10}=0.010 h^{-1}
  \Mpc$  (M6d), $R_{10}=0.015  h^{-1} \Mpc$  (M6e),  and $R_{10}=0.020
  h^{-1} \Mpc$ (M6f). See text for detailed discussion.}
\label{fig:radprof}
\end{figure*}

\section{The radial distribution of Satellite Galaxies}
\label{sec:rad}

The  hatched  histograms   in  Fig.~\ref{fig:radprof}  show  the  {\it
  projected}, radial,  number density distribution, $\bar{\Sigma}_{\rm
  sat}(R_p)$  (in $h^2/{\rm  Mpc}^2/{\rm host}$)  of  2dFGRS satellite
galaxies  around hosts in  three different  host magnitude  bins. Note
that  the $\bar{\Sigma}_{\rm sat}(R_p)$  reveal both  an inner  and an
outer break.   The outer break  radius increases with  increasing host
luminosity, and  is mainly due  to our selection criteria.   The inner
break  radius, however,  indicates a  pronounced absence  of satellite
galaxies at  small projected radii from  the host galaxy.  In order to
interpret  these findings  in  terms of  the actual  three-dimensional
number  density distribution  of satellite  galaxies we  compare these
findings to those obtained from our MGRSs.

The   three   non-hatched   histograms   in  the   upper   panels   of
Fig.~\ref{fig:radprof}  depict  the  $\bar{\Sigma}_{\rm sat}(R_p)$  as
obtained  from MGRSs  M1 (dotted  lines),  M6 (solid  lines), and  M12
(dashed lines).   Note that all three MGRSs  yield virtually identical
radial  number density  distributions.  In  fact, this  holds  for all
twelve MGRSs listed  in Table~1: for clarity, however,  only three are
shown.   Although the  projected number  density distributions  of the
MGRSs  nicely match  the 2dFGRS  data at  large projected  radii, they
severely  overpredict $\bar{\Sigma}_{\rm  sat}(R_p)$ at  small  $R_p$. 
Rather  than  a  pronounced  inner  break  radius,  $\bar{\Sigma}_{\rm
  sat}(R_p)$ increases continuously down  to the smallest radii shown. 
Especially  for the  brightest host  galaxies the  discrepancy between
MGRSs and 2dFGRS is dramatic.  Either satellite galaxies are spatially
anti-biased with respect to the dark matter, or the 2dFGRS has somehow
missed a large  number of satellite galaxies at  small projected radii
from their host galaxies. Below we investigate both options.

\subsection{Spatial anti-bias}
\label{sec:antibias}

When constructing  the MGRSs we  have thus far assumed  that satellite
galaxies follow  the same number  density distribution as  dark matter
particles (i.e., eq~[\ref{nsatr}] with $\alpha = {\cal R} = 1.0$). The
resulting overestimate  of the  projected number density  of satellite
galaxies at small  $R_p$, however, might indicate that  $\alpha < 1.0$
and/or that  ${\cal R} >  1$. To test  these ideas we  construct three
MGRSs using  CLF 6.  These  MGRSs, termed M6a,  M6b and M6c,  all have
$\alpha = 0.0$,  and only differ in the value of  ${\cal R}$, which is
set  to $1.0$, $2.0$,  and $3.0$,  respectively. The  projected number
density  distributions of their  satellite galaxies  are shown  in the
middle row of panels of Fig.~\ref{fig:radprof}. For comparison we also
plot  the  $\bar{\Sigma}_{\rm  sat}(R_p)$  of M6  (solid  lines).   As
expected, lowering $\alpha$ and increasing ${\cal R}$ both result in a
decrease of $\bar{\Sigma}_{\rm sat}$ at small $R_p$. In fact, MGRS M6c
fits most  of the  2dFGRS data reasonably  well, although it  fails to
reproduce the pronounced inner break evident in the data.

Taking these  results at face  value seems to indicate  that satellite
galaxies  are spatially anti-biased  with respect  to the  dark matter
mass  distribution on small  scales (20  to 30  percent of  the virial
radius).   This  is  in   wonderful  agreement  with  high  resolution
numerical $N$-body  simulations, which  reveal a very  similar spatial
anti-bias for  dark matter subhaloes  (Ghigna \etal 1998,  2000; Colin
\etal 1999; Okamoto \& Habe  1999; Springel \etal 2001; De Lucia \etal
2004; Diemand  \etal 2004), but inconsistent with  observations of the
number density  distribution of galaxies  in clusters (Beers  \& Tonry
1986; Carlberg  \etal 1997;  van der Marel  \etal 2000;  Diemand \etal
2004;   Lin,  Mohr  \&   Stanford  2004)   and  with   predictions  of
semi-analytical  models  of  galaxy  formation (Springel  \etal  2001;
Diaferio \etal 2001; Gao \etal 2004b).

\subsection{Observational projection effects}
\label{sec:project}

This  puzzling inconsistency  calls for  a closer  examination  of the
selection  effects in  the  2dFGRS.  Could  the  absence of  satellite
galaxies at small  projected radii from their host  galaxies be due to
selection effects?   Note that it can  not be explained as  due to the
fiber-collisions.  As shown in Fig.~\ref{fig:rtheta}, and discussed in
Section~\ref{sec:mock}, these effects  are accurately accounted for in
our MGRSs.  One  effect that has not yet  been accounted for, however,
is related to the fact that  galaxies have a finite size, which causes
them  to  overlap  in  projection.   Whenever this  happens  only  the
brightest of  the two will be  recognized as a galaxy  and included in
the survey.  Indeed, as shown  by Cole \etal (2001) using a comparison
of the  2dFGRS with the Two  Micron All Sky Survey  (2MASS), about 4.5
percent  of all  galaxies is  missed  in the  2dFGRS parent  catalogue
because  of merged  or  close  images (explaining  about  half of  the
incompleteness;  the other  half  being due  to incorrect  star-galaxy
separation)

In order to take this size-projection effect into account we proceed as
follows.  We model the characteristic size of a galaxy as
\begin{equation}
\label{size}
R_{\rm gal} = R_{10} \, \left( {L \over 10^{10} h^{-2} \Lsun} \right)^{\zeta}
\end{equation}
and define  the critical projection  angle $\theta_{\rm max}  = R_{\rm
  gal} / D_A$,  with $D_A$ the angular distance  of the galaxy.  Based
on  the  data on  disk  galaxies in  Courteau  \etal  (2004) we  adopt
$\zeta=1/3$, but we  note that this assumption is  not very important. 
For example,  we have also experimented with  $\zeta=1/2$ finding very
similar results as those described here.

We construct three  MGRSs in which we take  the size-projection effect
into account using $R_{10} = 0.010 h^{-1} \Mpc$ (M6d), $R_{10} = 0.015
h^{-1} \Mpc$ (M6e), and $R_{10} = 0.020 h^{-1} \Mpc$ (M6f).  We follow
the same procedure as described in Section~\ref{sec:mock}, except that
this   time,   after   having   corrected  for   the   position-   and
magnitude-dependent incompleteness, we compute the angular separations
$\theta$ between all galaxy pairs  and remove the faintest galaxy from
all  pairs  for which  $\theta  <  \theta_{\rm  max}$. Typically  this
removes about 2  to 4 percent of all galaxies,  which is comparable to
the incompleteness  level in the  2dFGRS parent catalogue due  to this
effect as found  by Cole \etal (2001).  Next we  correct for the fiber
collisions,  using the  same  trial-and-error method  as described  in
Section~\ref{sec:mock}.   Finally  we  remove  a  number  of  galaxies
completely at random  to bring the total fraction  of removed galaxies
to 9 percent.

The  lower panels  Fig.~\ref{fig:radprof} plot  the $\bar{\Sigma}_{\rm
  sat}(R_p)$ of MGRSs M6d  (dashed lines), M6e (dot-dashed lines), and
M6f (dotted lines).   Note that these MGRSs have $\alpha  = {\cal R} =
1.0$, as for our fiducial  MGRSs. For comparison, the solid lines plot
the projected  number density distributions  of MGRS M6, for  which no
correction for the size-projection  effect has been applied.  Clearly,
the size-projection  effect as modeled  here has a dramatic  impact on
$\bar{\Sigma}_{\rm sat}(R_p)$  at small  projected radii. In  fact, it
reproduces the inner cut-off  observed for the 2dFGRS data, indicating
that,  as first  noticed  by Cole  \etal  (2001), the  2dFGRS lacks  a
significant fraction of close galaxy pairs due to projection effects.

Unfortunately, unless  we can model this  particular incompleteness in
more  detail  as done  here  (i.e.,  have  independent constraints  on
$R_{10}$ and $\zeta$), we can  not use the 2dFGRS data to meaningfully
constrain  the   actual  radial  density   distribution  of  satellite
galaxies.   Our models  suggest that  the data  is consistent  with no
spatial (anti)-bias  as long as the size-projection  effects are taken
into account, but  we can not rule out that  $\alpha= 0.0$ and/or that
${\cal  R}>1.0$.  For example,  models with  both $\alpha=0.0$  and in
which  we  correct  for   the  size-projection  effect  are  virtually
indistinguisahble  from the  same  models but  with $\alpha=1.0$.   It
remains to be  seen whether the higher spatial  resolution of the SDSS
will allow a more in-depth investigation of the radial distribution of
satellite galaxies.
\begin{figure*}
\centerline{\psfig{figure=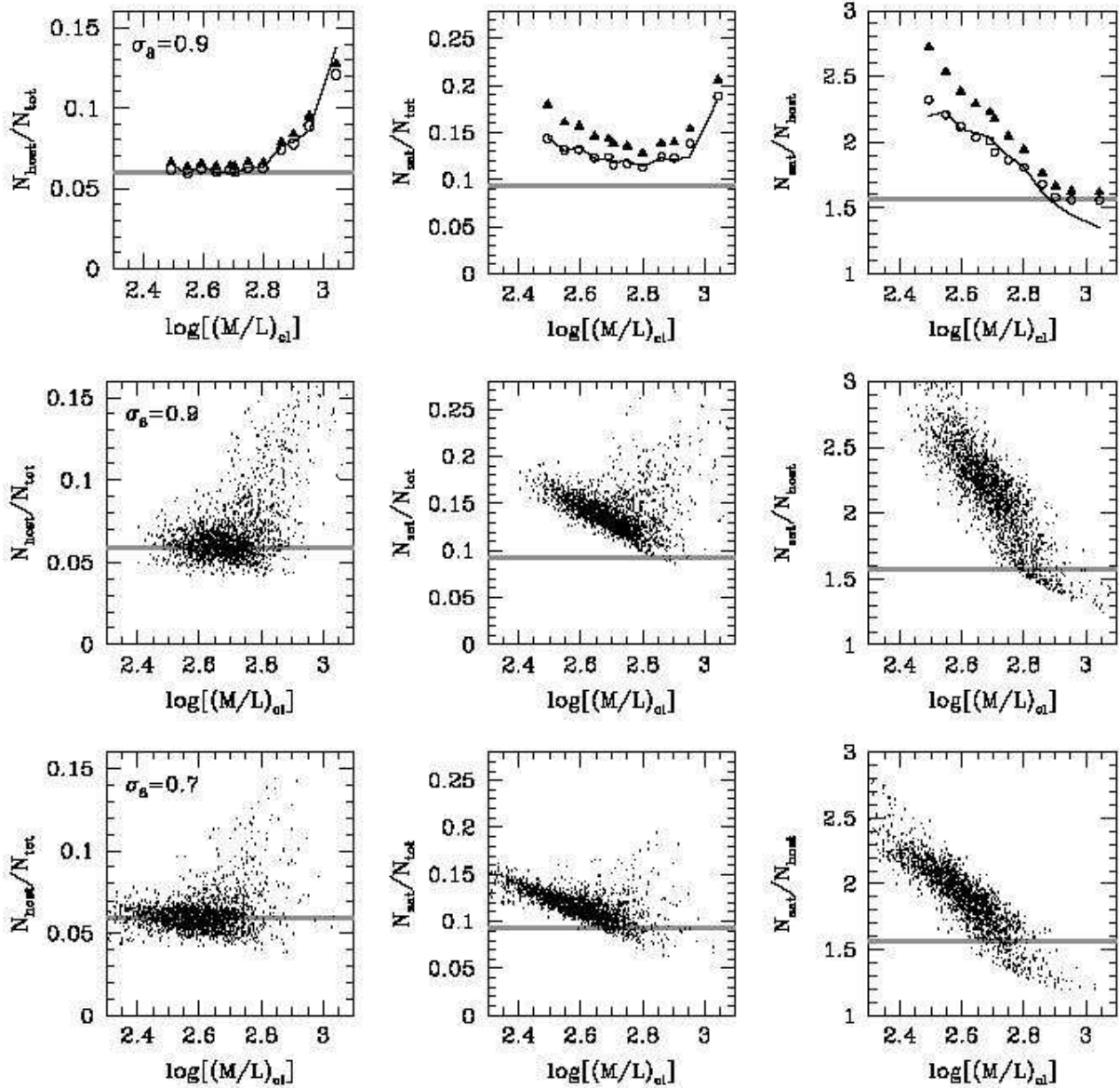,width=0.9\hdsize}}
\caption{Abundances of host and satellite galaxies. From left to
  right, the different  panels in a given row  plot the ratios $N_{\rm
    host}/N_{\rm   tot}$,  $N_{\rm   sat}/N_{\rm  tot}$   and  $N_{\rm
    sat}/N_{\rm host}$, all as function of cluster mass-to-light ratio
  $(M/L)_{\rm  cl}$ (in units  of $h  \MLsun$). Gray,  horizontal bars
  indicate the  ratios obtained from  the 2dFGRS. {\it  Upper panels:}
  solid triangles  and open circles indicate the  ratios obtained from
  the MGRSs  for the  twelve CLFs listed  in Table~1 without  and with
  correction   for   the   size-projection  effects,   respectively.   
  Errorbars,  computed assuming  Poissonian errors  on  $N_{\rm tot}$,
  $N_{\rm host}$ and $N_{\rm sat}$,  are smaller than the symbols, and
  are  therefore  not shown.   The  solid  line  indicates the  ratios
  obtained         from        the         analytical        estimates
  (eq.~[\ref{ntot}]--[\ref{nhostaver}])  after taking  account  of the
  correction factors $f_{\rm host}$  and $f_{\rm sat}$ (see text). The
  good agreement with the open circles indicates that these analytical
  estimates  can  be  used  to  compute the  abundances  of  host  and
  satellite galaxies for any CLF model without the need to construct a
  detailed MGRS. {\it Middle  panels:} the ratios $N_{\rm host}/N_{\rm
    tot}$,  $N_{\rm sat}/N_{\rm tot}$  and $N_{\rm  sat}/N_{\rm host}$
  for all  2000 samples in the  $\sigma_8 = 0.9$  MCMC, obtained using
  the approximate,  analytical method. Note that none  of these models
  can simultaneously match all three  ratios obtained from the 2dFGRS. 
  {\it Lower panels:} Same as middle panels, but this time showing the
  results for  the $\sigma_8 = 0.7$  MCMC.  Note that  at around ${\rm
    log}[(M/L)_{\rm   cl}]  \simeq   2.75$  these   model  predictions
  simultaneously match all three 2dFGRS ratios.}
\label{fig:abundances}
\end{figure*}

\section{The Abundance of Host and Satellite Galaxies}
\label{sec:abun}

We now  focus on the abundances  of host and  satellite galaxies.  For
the  2dFGRS  we  obtain   host  and  satellite  fractions  of  $N_{\rm
  host}/N_{\rm tot}  = 0.060$ $(0.060,0.059)$  and $N_{\rm sat}/N_{\rm
  host}  =  0.094$  $(0.096,0.092)$,  where the  numbers  in  brackets
indicate  the fractions  obtained using  only  the NGP  and SGP  data,
respectively.  We  compare these numbers with those  obtained from the
MGRSs.     The   solid    triangles   in    the   upper    panels   of
Fig.~\ref{fig:abundances}  plot  $N_{\rm  host}/N_{\rm tot}$,  $N_{\rm
  sat}/N_{\rm  tot}$, and  $N_{\rm sat}/N_{\rm  host}$ as  function of
$(M/L)_{\rm cl}$  for the twelve  MGRSs listed in Table~1.   The grey,
horizontal bars indicate the  2dFGRS results. Clearly, the fraction of
satellite  galaxies in  the  MGRS is  much  too high  compared to  the
2dFGRS,  independent  of  $(M/L)_{\rm  cl}$.   The  fraction  of  host
galaxies is  in reasonable agreement with the  2dFGRS (though somewhat
too high), but  only for $(M/L)_{\rm cl} \lta  650 h \MLsun$. Matching
the number of satellites per  host, however, favors much higher values
for  $(M/L)_{\rm  cl}$.   In  short,  none  of  the  twelve  MGRS  can
simultaneously match the abundances  of host and satellite galaxies in
the 2dFGRS.

The  open circles  in  the upper  panels of  Fig.~\ref{fig:abundances}
correspond to the  same twelve MGRSs, but this  time corrected for the
size-projection  effects discussed in  Section~\ref{sec:project} using
$R_{10}  = 0.015  h^{-1} \Mpc$  and $\zeta  = 1/3$.   This  lowers the
overall abundances  of both hosts and (mainly)  satellites, but still,
none of  the twelve  MGRS can simultaneously  match the  abundances of
host and satellite galaxies in  the 2dFGRS. Although the host fraction
is in  good agreement  with the  data for $(M/L)_{\rm  cl} \lta  650 h
\MLsun$,  matching   the  number  of  satellites   per  host  requires
$(M/L)_{\rm cl} \gta 800 h \MLsun$.

In order to investigate this failure of the models in more detail, and
to  explore all  freedom  in  the CLF  parameters,  ideally one  would
construct  a MGRS  for each  of the  2000 models  in our  MCMC.  This,
however,  is not  feasible computationally,  and we  therefore  use an
approximate, analytical method instead.   The total number of galaxies
in a flux-limited survey is given by
\begin{equation}
\label{ntot}
N_{\rm tot} = \int\limits_{0}^{\Omega} {\rm d}\Omega
\int\limits_{z_{\rm min}}^{z_{\rm max}} {\rm d}z \int\limits_0^{\infty} {\rm d}M 
\, n(M) \, \langle N \rangle_{M,z}
\end{equation}
Here $\Omega$ is  the solid angle of sky of the  survey, ${\rm d}V$ is
the differential  volume element, $z_{\rm min}$ and  $z_{\rm max}$ are
the  survey redshift  limits ($z_{\rm  min}=0.01$ and  $z_{\rm  max} =
0.15$  in our  case), and  $\langle  N \rangle_{M,z}$  is the  average
number of galaxies per halo of  mass $M$ at redshift $z$ which follows
from the CLF according to
\begin{equation}
\label{naver}
\langle N \rangle_{M,z} = \int_{L_{\rm min}(z)}^{L_{\rm max}(z)} 
\Phi(L \vert M) \, {\rm d}L 
\end{equation}
with  $L_{\rm min}(z)$ and  $L_{\rm max}(z)$  the minimum  and maximum
luminosities  of a  galaxy at  redshift  $z$ that  makes the  apparent
magnitude limits of  the survey (we adopt $15.0 <  m_{b_J} < 19.3$ for
the 2dFGRS).  Similarly, the number of satellite galaxies follows from
\begin{equation}
\label{nsat}
N_{\rm sat} = \int\limits_{0}^{\Omega} {\rm d}\Omega
\int\limits_{z_{\rm min}}^{z_{\rm max}} {\rm d}z \int\limits_0^{\infty} {\rm d}M 
\, n(M) \, \langle N_{\rm sat} \rangle_{M,z}
\end{equation}
with
\begin{equation}
\label{nsataver}
\langle N_{\rm sat} \rangle_{M,z} = \left\{ \begin{array}{ll}
\langle N \rangle_{M,z} - 1 & \mbox{if $\langle N \rangle_{M,z} \geq 1$} \\
0                          & \mbox{otherwise}
\end{array} \right.
\end{equation}
The  number of  host galaxies  in a  flux-limited survey,  finally, is
given by
\begin{equation}
\label{nhost}
N_{\rm host} = \int\limits_{0}^{\Omega} {\rm d}\Omega
\int\limits_{z_{\rm min}}^{z_{\rm max}} {\rm d}z \int\limits_0^{\infty} {\rm d}M
\, n(M) \, w(M,z)
\end{equation}
where the weight function $w(M,z)$ is given by
\begin{equation}
\label{nhostaver}
w(M,z) = \left\{ \begin{array}{ll}
1 - {\rm exp}\left( -\langle N_{\rm sat} \rangle_{M,z} \right) & \mbox{if $\langle N \rangle_{M,z} \geq 1$} \\
0   & \mbox{otherwise}
\end{array} \right.
\end{equation}
This derives from the fact that only hosts with at least one satellite
galaxy  are counted  as host  galaxies, and  using the  fact  that the
number of  satellite galaxies  follows a Poissonian  distribution (see
van den Bosch \etal 2004b).

Using these equations we compute  the abundances of host and satellite
galaxies for each  of the 2000 CLF models in  the MCMC.  However, this
does not take care of incompleteness effects, fiber collisions, or the
size-projection effect.  In addition, this assumes that all satellites
are  selected  with  zero  interlopers.  Instead,  the  host-satellite
selection  criterion used  for the  2dFGRS and  the MGRSs  is  not 100
percent  complete and  yields about  15 percent  interlopers  (van den
Bosch \etal 2004b).  To  correct for all these effects/shortcomings we
proceed as follows.  We multiply  the $N_{\rm host}$ and $N_{\rm sat}$
computed  using  eq.~(\ref{nhost})  and~(\ref{nsat})  with  correction
factors  $f_{\rm  host}$ and  $f_{\rm  sat}$,  respectively, which  we
calibrate using the results of  the 12 MGRSs (that have been corrected
for  the size-projection  effects) shown  in Fig.~\ref{fig:abundances}
(open circles).   We find a  good match for  $f_{\rm host} =  0.8$ and
$f_{\rm  sat} =  0.57  +  0.25[{\rm log}(M/L)_{\rm  cl}  - 3.0]$,  the
results of which  are indicated by the solid line  in the upper panels
of Fig.~\ref{fig:abundances}.  In what follows we assume that the same
correction factors apply to all models in our MCMC.

Although    only    approximate,   eq.~(\ref{ntot})--(\ref{nhostaver})
combined with  this simple scaling  allows us to make  predictions for
the fractions of host and satellite  galaxies for each of the 2000 CLF
models  in our  MGRSs.  The  results are  shown in  the middle  row of
panels of Fig.~\ref{fig:abundances}.   The overall behavior is similar
to  that of  the 12  MGRSs  shown in  the upper  panels: matching  the
fraction of host galaxies requires $(M/L)_{\rm cl} \lta 650 h \MLsun$,
whereas matching the number  of satellites per host basically requires
the  opposite.   Meanwhile,  the   number  of  satellite  galaxies  is
systematically too  high. Although there  are a few models  that match
the fraction  of satellite galaxies of  the 2dFGRS, as  we show below,
these  models  can  not  simultaneously  match the  fraction  of  host
galaxies. Therefore, we conclude that  {\it the abundances of host and
  satellite galaxies  in the 2dFGRS  can not be reproduced  within the
  $\Lambda$CDM concordance cosmology with $\sigma_8=0.9$}.

\subsection{Evidence for a low value of $\sigma_8$}
\label{sec:sigma}

Using MGRSs similar  to those presented here, Yang  \etal (2004a) have
shown that a $\Lambda$CDM concordance cosmology with $(M/L)_{\rm cl} =
500 h \MLsun$ predicts too  much clustering power on small scales, and
pairwise  peculiar velocity dispersions  that are  too high.   In Yang
\etal (2004b) it was shown that the same model predicts too many large
groups of  galaxies. All  these problems indicate  that there  are too
many  galaxies in  (massive)  clusters compared  to observations.   As
demonstrated in Yang \etal (2004a,b), this can be remedied by adopting
a high $(M/L)_{\rm cl}$, which  results in fewer galaxies per cluster. 
This  agrees perfectly  with  the  fact that  matching  the number  of
satellites  per  host requires  a  similarly  high  $(M/L)_{\rm cl}$.  
However, the  failure to simultaneously match  the separate abundances
of  host and  satellite  galaxies, and  the independent  observational
constraints  on  cluster  mass-to-light  ratios  which  indicate  that
$(M/L)_{\rm cl} \simeq  (400 \pm 100) h \MLsun$  (Carlberg \etal 1996;
Fukugita, Hogan \&  Peebles 1998; Bahcall \etal 2000),  signal a clear
shortcoming of this model.
\begin{figure*}
\centerline{\psfig{figure=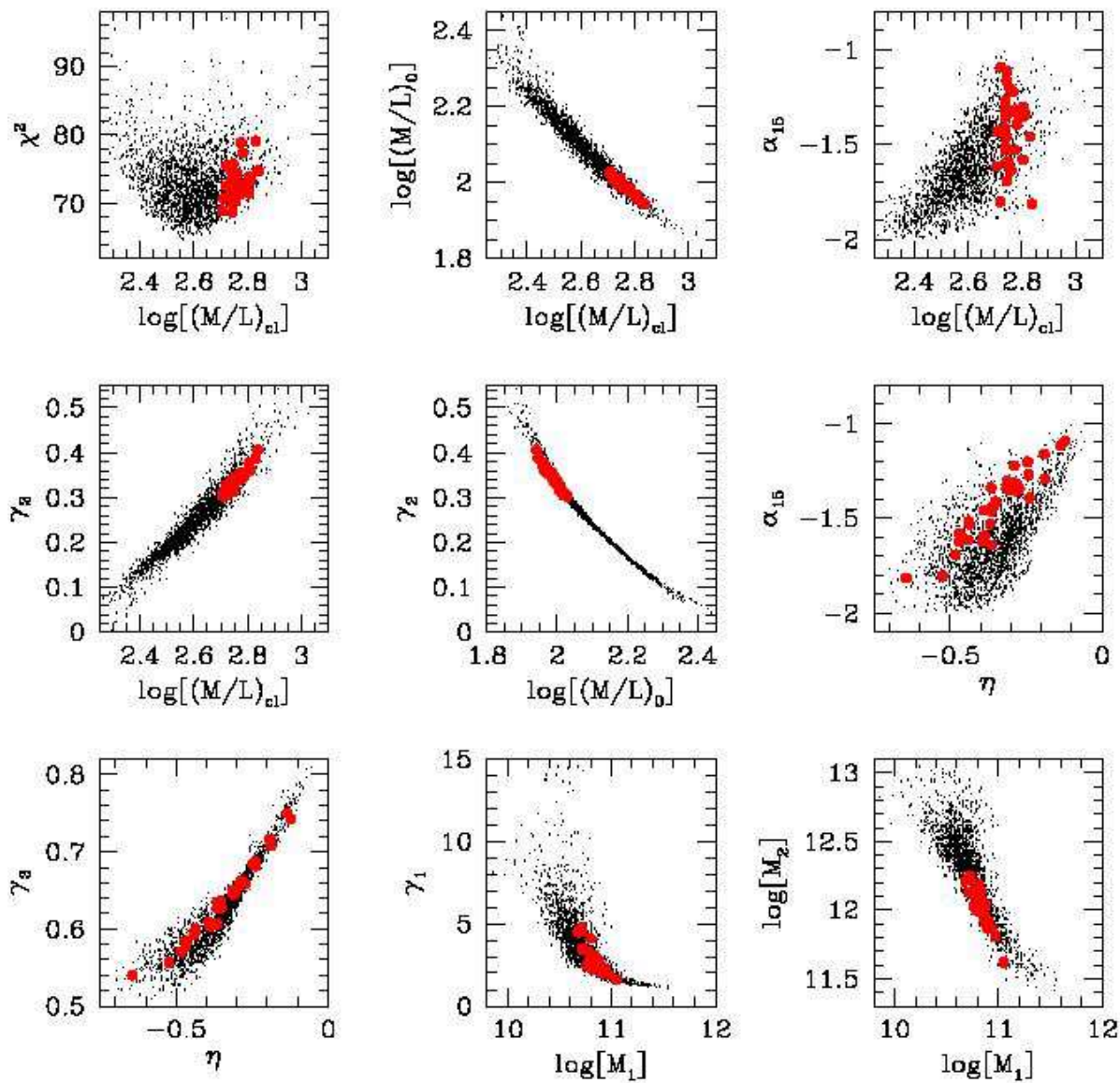,width=0.9\hdsize}}
\caption{Same as Fig.~\ref{fig:scatter} except that this time we plot 
  parameter   correlations  for  the   MCMC  constructed   assuming  a
  $\Lambda$CDM  cosmology with  $\sigma_8=0.7$. The  thick  solid dots
  indicate those models for which  $\chi^2_{\rm ab} < 4$, indicating a
  good  match  to  the  observed  abundances  of  host  and  satellite
  galaxies.   Note that  these extra  constraints severely  restrict a
  number  of  the  CLF  parameters;  cf.,  the  parameters  of  models
  $\Lambda_{0.7}$ and $\Lambda_{0.7}^{\rm ab}$ in Table~1.}
\label{fig:morescatter}
\end{figure*}

An  alternative  solution to  the  apparent  overabundance of  cluster
galaxies is to  lower the abundance of cluster-sized  haloes.  This is
most easily  established by lowering  the power-spectrum normalization
$\sigma_8$. As shown by Yang \etal (2004a,b), a $\Lambda$CDM cosmology
with $\sigma_8=0.7$  and $(M/L)_{\rm cl}  = 500 h \MLsun$  can equally
well match the clustering  data, the pairwise peculiar velocities, and
the  abundances  of galaxy  groups  as  a  concordance cosmology  with
$\sigma_8=0.9$ and  $(M/L)_{\rm cl} =  900 h \MLsun$. In  addition, as
shown  in van  den  Bosch \etal  (2003b),  this cosmology  is in  good
agreement with the WMAP data and may even alleviate some problems with
the $\Lambda$CDM cosmology regarding  the concentration of dark matter
haloes. It  is interesting, therefore,  to investigate whether  such a
model  can  also  simultaneously  match  the abundances  of  host  and
satellite galaxies in the 2dFGRS.

In  order  to test  this  we construct  a  MCMC  for the  $\Lambda$CDM
cosmology but with $\sigma_8=0.7$. We use the same number of steps and
the   same   thinning   factor   as   for  the   MCMC   described   in
Section~\ref{sec:mcm}.   The resulting  distribution of  parameters is
indicated by the  non-hatched histograms in Fig.~\ref{fig:histo}.  The
corresponding median and 68 percent confidence intervals are listed in
Table~1   (model   $\Lambda_{0.7}$),   Correlations  amongst   various
parameters are  shown in Fig.~\ref{fig:morescatter}.   Compared to the
distributions  for the concordance  cosmology with  $\sigma_8=0.9$ the
main  differences  are  a  reduction  of the  mean  $(M/L)_{\rm  cl}$,
$\alpha_{15}$  and  $\gamma_3$. Other  parameters,  notably $M_1$  and
$\gamma_1$, are extremely insensitive to $\sigma_8$.

Unfortunately we do not  have numerical simulations for a $\Lambda$CDM
cosmology with $\sigma_8=0.7$, so that  we can not construct MGRSs for
this cosmology  (but see Yang  \etal 2004a). Nevertheless, we  can use
the  scaling parameters $f_{\rm  host}$ and  $f_{\rm sat}$,  under the
assumption that they are also valid for this cosmology, to predict the
abundances  of  host and  satellite  galaxies  for the  low-$\sigma_8$
cosmology  using  eq.~(\ref{ntot})--(\ref{nhostaver}).  The  resulting
host-satellite   fractions  are   shown   in  the   lower  panels   of
Fig.~\ref{fig:abundances}.   Compared  to  the  same results  for  the
$\sigma_8=0.9$  cosmology  (panels   in  middle  row),  the  satellite
fractions have  been reduced, bringing  them in better  agreement with
the 2dFGRS results.   In order to make the  comparison between the two
different  $\sigma_8$  models more  quantitative,  and to  investigate
whether  any  model  can  simultaneously  match the abundances of host
and satellite galaxies,  we introduce the goodness-of-fit measure
\begin{equation}
\label{chiab}
\chi^2_{\rm ab} = 
\left({{N_{\rm host} \over N_{\rm tot}} - 0.060 \over 0.002}\right)^2 + 
\left({{N_{\rm sat}  \over N_{\rm tot}} - 0.094 \over 0.003}\right)^2 
\end{equation}
where the numbers  in the numerators are the  2dFGRS ratios, and those
in the denominators are the standard deviations due to cosmic variance
that  we  obtain from  a  set  of  independent MGRSs  using  different
simulation boxes  (see Yang \etal 2004a)\footnote{Note  that these are
  very similar to the  standard deviations obtainded from a comparison
  of the host/satellite fractions in the 2dFGRS NGP and SGP}.  For the
$\sigma_8 = 0.7$  cosmology, the number of samples (out  of a total of
2000) with  $\chi^2_{\rm  ab} <  (2,4,6)$  is  $(11,34,57)$.  For  the
$\sigma_8=0.9$ cosmology, these  numbers are $(0,0,0)$.  Clearly, {\it
  the $\sigma_8=0.7$  models are far more  succesful in simultaneously
  fitting  the abundances of  host and  satellite galaxies  than those
  with  $\sigma_8=0.9$.}   

The thick,  solid dots  in Fig.~\ref{fig:morescatter} indicate  the 34
CLF models in the  $\sigma_8=0.7$ cosmology for which $\chi^2_{\rm ab}
< 4$. Note  how they are clustered together  in parameter spaces. This
means  that  using  the  observed  abundances of  host  and  satellite
galaxies  in the  2dFGRS  as extra  constraints  allows a  significant
tightening  of the  constraints on  the CLF  parameters.  We  can take
account of  these additional constraints  by weighting each  sample in
the MCMC with ${\rm exp}(-\chi^2_{\rm ab}/2)$.  This yields the median
and   68    percent   confidence   intervals    listed   under   model
$\Lambda_{0.7}^{\rm  ab}$ in  Table~1.  Note  that the  constraints on
some of  the CLF parameters,  notably $(M/L)_0$, $(M/L)_{\rm  cl}$ and
$M_1$, are now much tighter (cf.  model $\Lambda_{0.7}$).

Although  the above  results strongly  favor a  $\Lambda$CDM cosmology
with relatively low power spectrum normalization, we caution that they
are obtained  assuming that the scaling parameters  $f_{\rm host}$ and
$f_{\rm sat}$, determined for a  small subset of all MCMC samples with
$\sigma_8=0.9$,  are  valid for  all  MCMC  samples,  even those  with
$\sigma_8=0.7$.   Testing  the accuracy  of  this assumption  requires
similar  $N$-body  simulations  as   those  used  here,  but  for  the
$\Lambda$CDM cosmology with $\sigma_8=0.7$,  and the construction of a
large number of different MGRSs to probe the full CLF parameter space.
Given our limited computational resources, we are unfortunately unable
to perform these tests.

\section{Summary}
\label{sec:summ}

Comparing  the  radial  distribution  and abundances  of  dark  matter
subhaloes in  simulations with  those of observed  satellite galaxies,
which are  thought to be associated  with these subhaloes,  has led to
two   apparent  inconsistencies:   (i)  The   radial   number  density
distribution  of subhaloes  reveals a  constant density  core, whereas
galaxies in clusters  seem to follow an NFW  profile, and (ii) whereas
dark matter haloes  of different masses look homologous  when it comes
to the  properties of  their dark matter  subhaloes, a  galaxy cluster
looks very different from a galaxy sized system when it comes to their
satellite galaxies (see Section~\ref{sec:intro} for references).

In this paper we have  made use of the conditional luminosity function
(CLF)  to  address  these  issues  using data  from  the  2dFGRS.   We
constructed large Monte Carlo Markov Chains that allowed us to explore
the  full posterior  distribution of  the CLF  parameter  space. Using
detailed mock galaxy redshift surveys (MGRSs) based on the CLF we have
analyzed the radial distribution  and abundances of host and satellite
galaxies in  the 2dFGRS.   Our main conclusions  can be  summarized as
follows
\begin{enumerate}
  
\item  The 2dFGRS  is missing  a significant  fraction of  galaxies in
  close (projected) pairs (about 2 to 4 percent of {\it all} galaxies)
  due to the overlap and merging of galaxy images in the APM catalogue
  (see also Cole \etal 2001).
  
\item  Due to  this close-pair  incompleteness we  can not  put strong
  constraints on the radial  distribution of satellite galaxies. After
  modelling the  close-pair incompleteness in  our MGRSs the  data are
  consistent  with  a  model   in  which  the  radial  number  density
  distribution of  satellite galaxies follows that of  the dark matter
  particles  (i.e., a  NFW  distribution),  but we  can  not rule  out
  alternatives with, for example, a constant number density core.
  
\item   Within    the   $\Lambda$CDM   concordance    cosmology   with
  $\sigma_8=0.9$  we can  not simulteneously  match the  abundances of
  host and  satellite galaxies in  the 2dFGRS. Matching the  number of
  satellites {\it per}  host requires exceptionally high mass-to-light
  ratios  on cluster  scales.   This is  in  excellent agreement  with
  previous  findings   based  on  the  CLF   formalism,  but  severely
  overpredicts the abundance of both host and satellite galaxies.
  
\item Simultaneously matching  the luminosity function, the clustering
  properties as function of luminosity, and the abundances of host and
  satellite  galaxies  seems  to  require  a reduction  of  the  power
  spectrum normalization  to $\sigma_8 \simeq 0.7$.  As  shown by Yang
  \etal  (2004a,b),  this is  also  required  in  order to  match  the
  pairwise peculiar velocities in  the 2dFGRS, the clustering power on
  small scales, and the multiplicity function of galaxy groups.
  
\item The  CLF models for the $\Lambda$CDM  concordance cosmology with
  $\sigma_8=0.7$  that  match the  abundances  of  host and  satellite
  galaxies  are extremely  well  constrained. They  indicate that  the
  average  mass-to-light  ratio  of   dark  matter  haloes  reveals  a
  pronounced minimum  of $97 \pm 3 h  \MLsun$ at a halo  mass of $(7.2
  \pm 0.7)  \times 10^{10} h^{-1}  \Msun$. On cluster scales  ($M \gta
  10^{14} h^{-1}  \Msun$) the average mass-to-light ratio  is $570 \pm
  40  h \MLsun$.  
 
\end{enumerate}
%


\section*{Acknowledgements}

We are grateful  to Yipeng Jing for providing us  the set of numerical
simulations  used for  the construction  of our  mock  galaxy redshift
surveys, and to Ed Hawkins  for providing his fitting function used to
model fiber collisions.  Ben  Moore and Juerg Diemand are acknowledged
for  detailed  discussions.  PN   acknowledges  receipt  of  a  Zwicky
fellowship.



\label{lastpage}

\end{document}